\documentclass[lettersize,journal]{IEEEtran}
\usepackage{amsmath,amsfonts}
\usepackage{array}
\usepackage{cite}
\usepackage[caption=false,font=normalsize,labelfont=sf,textfont=sf]{subfig}
\usepackage{textcomp}
\usepackage{stfloats}
\usepackage{pifont}
\usepackage{verbatim}
\usepackage{amssymb}
\usepackage[utf8]{inputenc}   % 仅 pdflatex 需要
\usepackage[T1]{fontenc}
\usepackage{amsmath}
\usepackage{algorithm}
\usepackage{booktabs} % 用于更好的表格线条  
\usepackage{siunitx}  % 用于列对齐和数学符号  
\usepackage{algpseudocode}
\usepackage{graphicx}
\usepackage{multirow}  % 用于表格多行合并
\usepackage{bm}
\usepackage{booktabs} % 确保导言区引入了 booktabs 宏包
\usepackage{multirow}
\usepackage{array}
\usepackage{booktabs}
\usepackage{pifont}   % for \ding{51} checkmark
\newcommand{\cmark}{\ding{51}}
\def\BibTeX{{\rm B\kern-.05em{\sc i\kern-.025em b}\kern-.08em
		T\kern-.1667em\lower.7ex\hbox{E}\kern-.125emX}}
\usepackage{balance}

\begin{document}
	\title{Deep Mixture of Experts Network for Resource Optimization in
		Aerial-Terrestrial CF-mMIMO Systems under URLLC}
	
	\author{Donggen Li, $\textit{Graduate Student Member, IEEE}$, Chong Huang, $\textit{Member, IEEE}$, Jingfu Li, $\textit{Member, IEEE}$,\\ Pei Xiao, $\textit{Senior Member, IEEE}$, Wenjiang Feng, Dusit Niyato, $\textit{Fellow, IEEE}$, Zhu Han, $\textit{Fellow, IEEE}$ 
		
	\thanks{
	\par D. Li, W. Feng are with the School of Microelectronics and Communication Engineering, Chongqing University, Chongqing 400044, China (email: lidonggen@stu.cqu.edu.cn; fengwj@cqu.edu.cn).
	\par C. Huang and P. Xiao are with the 5GIC and 6GIC, Institute for Communication Systems, University of Surrey, Guildford GU2 7XH, U.K. (email: chong.huang@surrey.ac.uk; p.xiao@surrey.ac.uk).
	
	\par J. Li is with the School of Information Science and Technology, Southwest Jiaotong University, Chengdu, 611756, China. (email: jingfuli@swjtu.edu.cn).
	
	\par D. Niyato is with the School of Computer Science and Engineering, Nanyang Technological University, Singapore 639798 (email: DNIYATO@ntu.edu.sg).

	\par Z. Han is with the Department of Electrical and Computer Engineering at the University of Houston, Houston, TX 77004 USA, and also with the Department of Computer Science and Engineering, Kyung Hee University, Seoul, South Korea, 446-701 (email: hanzhu22@gmail.com).
	}	
}
\maketitle

	\begin{abstract}
		
	As a critical component of sixth-generation (6G) wireless networks, ultra-reliable and low-latency communication (URLLC) is expected to support real-time and reliable information exchange in low-altitude environments. However, achieving URLLC often incurs significant resource overhead, including increased bandwidth consumption, higher transmit power, and denser access point (AP) deployment, which pose significant challenges to both spectral efficiency (SE) and energy efficiency (EE). Besides, existing iterative optimization algorithms are computationally intensive and struggle to meet the latency requirements of URLLC. To address these challenges, we propose a hybrid aerial-terrestrial cell-free massive MIMO (CF-mMIMO) network to support diverse services, along with a channel prediction network and a deep mixture of experts (MoE) network for uplink optimization. First, we design a channel prediction network (CP-Net) to mitigate channel aging caused by high-mobility user equipment (UE). CP-Net employs three Transformer-based sub-networks for aged channel state information (CSI) prediction, while a channel quality-aware loss function is introduced to improve the prediction accuracy of weak links. Based on the predicted CSI, we develop a deep MoE network (MoE-Net) for power allocation comprising three expert models targeting different objectives. Then, we introduce a weighted gating network (WT-Net) to learn an efficient adaptive combination of expert outputs. The proposed framework better captures heterogeneous UE requirements and improves communication performance under URLLC constraints. Numerical results demonstrate the effectiveness of the proposed method.

	\end{abstract}


	\section{Introduction}
	\subsection{Background}
	%%%%开头，URLLC->low-atitude.
	
	\par Ultra-reliable and low-latency communication (URLLC), as a critical technology in sixth-generation (6G) mobile communication systems, is designed to support mission-critical data transmission with stringent quality-of-service (QoS) requirements \cite{10507759,8705373,10535307}. This technology has found extensive applications in emerging low-altitude scenarios, such as urban air mobility (UAM), low-altitude smart logistics (LASL), and unmanned traffic management (UTM) \cite{10723207}. These scenarios involve high-density aerial operations over wide geographical areas, introducing considerable challenges such as maintaining channel coherence, enhancing Doppler resilience, and managing inter-user interference (IUI) \cite{8705373}. In conventional cellular massive MIMO networks, user equipment (UE) connects to the nearest access point (AP), which induces severe cell-edge interference and prevents reliable communication \cite{10411070}. In contrast, cell-free massive MIMO (CF-mMIMO) employs a distributed AP architecture to mitigate interference for edge UEs and reduce the distance between UEs and APs \cite{elhoushy2021cell,8630677,10461045}. In this architecture, signals are received by multiple APs and transmitted to a central processing unit (CPU) via fronthaul links for joint signal processing, making CF-mMIMO particularly suitable for supporting URLLC services \cite{10461045}.

	\par To satisfy the stringent URLLC requirements in future wireless networks, higher transmit power, wider bandwidth, and larger antenna configurations are typically required, resulting in significant spectral and energy overheads in CF-mMIMO networks \cite{10535307}. Therefore, improving spectral efficiency (SE) and energy efficiency (EE) through dynamic resource management is essential for realizing URLLC-enabled systems \cite{10689367}. To address these issues, iterative resource optimization algorithms have been widely studied, generally employing two strategies: exploring the solution space to approximate an optimal solution, or reformulating non-convex problems into convex ones for efficient optimization \cite{11018229}. However, these methods often rely on repeated iterative updates and may incur non-negligible, input-dependent decision overhead, which makes fast adaptation challenging in highly dynamic URLLC-oriented scenarios. Moreover, in communication scenarios with heterogeneous service demands, conventional model-based designs often entail higher modeling complexity and may be less flexible under varying operating conditions \cite{10621640}.

    \par In low-altitude communication scenarios, obstacle blockages, scattering, reflection, and diffraction significantly affect signal propagation \cite{10723207}. Moreover, the high mobility of aerial devices induces rapid spatio-temporal channel variations, leading to channel aging, multipath interference, and frequency-selective fading, which challenge the realization of URLLC applications \cite{chopra2017performance}. In recent years, deep neural networks (DNNs) have attracted considerable attention in wireless resource optimization \cite{lecun2015deep}. By leveraging their ability to capture inherent structural features, DNNs can directly map channel states to near-optimal resource allocations, thereby avoiding computationally intensive iterative procedures and facilitating real-time resource management compared to traditional approaches \cite{10621640}. Furthermore, deep learning-based channel prediction methods can capture the spatio-temporal dynamics of wireless channels, providing more reliable channel state information (CSI). This facilitates the mitigation of time-varying channel impairments, thereby reducing interference and enhancing the reliability of URLLC service \cite{10716774}.

	\subsection{Related Work}
    To expand communication coverage and exploit spatial-diversity gains, unmanned aerial vehicle (UAV)-assisted CF-mMIMO architectures have been widely investigated across diverse scenarios \cite{10634184, 10547003, 11045816, 10494364}. In \cite{10547003}, an uplink aerial CF-mMIMO framework is studied, where aerial APs serve ground UEs and forward locally processed signals to a CPU-UAV for one-shot fusion under correlated Rician fading. The proposed distributed detection and combining design improves SE while reducing centralized complexity. In \cite{11045816}, a reconfigurable intelligent surface (RIS)-assisted downlink CF-mMIMO system with co-existing aerial UAV and ground users is investigated. The system employs a max–min signal-to-interference-plus-noise ratio (SINR) formulation to jointly optimize AP power allocation and RIS phase shifts. In \cite{10494364}, a multi-UAV-aided cell-free radio access network (RAN) with network-assisted full-duplex operation for URLLC is analyzed, where multiple UAV-APs serve ground users and the downlink URLLC performance is characterized using finite blocklength (FBL) transmission. However, existing studies rarely consider the joint presence of aerial APs and UEs \cite{10547003}, and URLLC-oriented architectural design for heterogeneous objectives has also received limited attention, which restricts their applicability to heterogeneous aerial mobility scenarios \cite{10494364}. Therefore, more efficient and comprehensive CF-mMIMO architectures still require further investigation.

    \par In low-altitude scenarios, the high mobility of aerial and ground UEs leads to severe channel aging, where outdated CSI significantly degrades the performance of deep learning--based resource-allocation methods \cite{10859261,9832933,9676455}. In \cite{chopra2017performance}, channel aging in frequency-division-duplex (FDD) massive MIMO systems was analyzed, and channel-estimation error bounds and SINR approximations were derived. It was shown that an optimal frame duration can mitigate channel aging effects in high-mobility scenarios. In \cite{10859261}, an extended Kalman-filter-based directional CSI prediction scheme with centralized hybrid precoding was proposed, demonstrating an effective trade-off between model complexity and prediction accuracy. In \cite{9832933}, a Transformer-based channel prediction architecture was proposed to exploit sequence modeling and long-range temporal dependencies, thereby improving average CSI-prediction accuracy. In \cite{9676455}, a spatial--temporal channel prediction framework for high-speed railway mMIMO was developed, where a propagation-graph-generated dataset was used to train a model that maps historical channel observations to future channel states. By jointly leveraging temporal dynamics and spatial correlations, it outperforms several classical and neural baselines. However, existing deep learning methods trained with aggregate objectives tend to favor high-quality channels, which are easier to fit due to their clearer structural patterns. This training bias can degrade prediction accuracy for low-quality channels near the URLLC boundary, thereby reducing the URLLC reliability satisfaction probability \cite{9832933}.

    \par Deep learning--based optimization approaches have been widely explored for URLLC in CF-mMIMO systems. Compared with conventional iterative methods, these approaches avoid costly alternating updates and exhibit stronger generalization in high-dimensional non-convex optimization problems \cite{10924175, 10045791,kim2024transformer}. In \cite{10924175}, an energy-efficient uplink design for CF-mMIMO systems was investigated. Specifically, a two-stage iterative power-allocation algorithm was proposed, and a supervised deep-learning framework was developed to approximate the resulting allocation policy, thereby enabling real-time resource allocation. In \cite{10045791}, unsupervised deep-learning-based uplink and downlink power-control strategies were studied, achieving near-optimal performance with low complexity in user-centric CF-mMIMO systems, where each UE is served by a subset of APs. In \cite{kim2024transformer}, a learning-based two-stage resource-allocation framework was proposed to support the coexistence of URLLC and enhanced mobile broadband (eMBB) in beyond-5G (B5G) networks, demonstrating improved trade-offs among fairness, throughput, and reliability compared with baseline methods. As communication scenarios become increasingly complex and service objectives more heterogeneous, it becomes difficult for a single unified network to simultaneously handle conflicting optimization priorities across different UE types. However, existing studies have paid limited attention to service partitioning and structured modeling tailored to such heterogeneous objectives \cite{10924175,10045791,kim2024transformer}.

   \begin{table*}[t]
    \centering
    \caption{COMPARISON BETWEEN OUR WORK AND EXISTING WORKS.}
    \label{tab:1}
    \setlength{\tabcolsep}{5pt}
    \renewcommand{\arraystretch}{1.05}
    \scriptsize
    
    \resizebox{\textwidth}{!}{%
    \begin{tabular}{|p{4.5cm}|c|*{9}{c|}}
    \hline
    Novelty & Our Work
    & [16] & [17] & [18]
    & [19] & [20]  & [21]
    & [22] & [23] & [24] \\
    \hline
    aerial AP deployment & \cmark& \cmark &  & \cmark&  &  &&  &  &  \\\hline
    Coexisting aerial and ground UEs & \cmark&  &  \cmark & &  &  & &  &  &  \\\hline
    URLLC constraints & \cmark&  &  &\cmark &  &  && \cmark &  &  \\\hline
    Channel aging & \cmark&  &  & &\cmark  & \cmark & \cmark&\cmark  &  & \cmark \\\hline
    Learning-based CSI prediction & \cmark&  &  & &  & \cmark & \cmark&  &  &  \\\hline
    Learning-based resource allocation & \cmark&  &  &&  &  && \cmark & \cmark & \cmark \\\hline
    Multiple link types& \cmark&  & \cmark &&  &  &&  &  &  \\\hline
    Weak-link-aware training & \cmark&  &  &&  &  &&  &  &  \\\hline
    Multi-tasks & \cmark&  &  &&  &  && \cmark &  &  \\\hline
    Heterogeneous user objectives & \cmark&  &  &  &  &  & &  &  &  \\\hline
    Expert-specialized resource allocation & \cmark&  & &&  &  &&  &  &  \\\hline
    \end{tabular}%
    }
\end{table*}

	\subsection{Contribution}
    Existing approaches face three main challenges: (i) existing CF-mMIMO architectures for aerial scenarios have paid limited attention to coexisting aerial and ground UEs, while URLLC-oriented designs for UAM remain underexplored \cite{10547003, 11045816, 10494364}, (ii) existing learning-based prediction methods tend to underemphasize weak links under aggregate-objective training, which can degrade the overall reliability performance \cite{10859261,9832933,9676455}, and (iii) under heterogeneous optimization requirements, a single unified model often struggles to generalize well across UEs with different objectives \cite{10924175,10045791,kim2024transformer}. This paper investigates the joint design of channel prediction and resource allocation for uplink optimization in low-altitude UAM scenarios with aged CSI, heterogeneous service requirements, and reliability and latency constraints for URLLC. Unlike existing studies, we develop a hybrid aerial--terrestrial CF-mMIMO framework for real-time information exchange, propose a link-quality-aware channel prediction network to mitigate channel-aging effects, and employ a mixture-of-experts (MoE) network \cite{10937907} for power allocation across specialized optimization objectives. Through adaptive decision fusion, the proposed framework is designed to improve scalability and robustness under dynamic channel conditions. Table~\ref{tab:1} presents an explicit comparison with existing studies, and the main contributions are summarized as follows:
	
	\begin{itemize}
		\item[$\bullet$] We introduce a hybrid aerial--terrestrial CF-mMIMO architecture into low-altitude UAM scenarios to support heterogeneous service objectives, where aerial and terrestrial APs jointly serve both aerial and ground UEs under centralized signal combining and resource allocation at CPU. We further incorporate FBL expressions to characterize short-packet transmission and formulate the corresponding URLLC constraints. 
		
		\item[$\bullet$] 
		We propose a channel prediction network (CP-Net) to enhance CSI accuracy under channel aging. Specifically, CP-Net employs three supervised Transformer-based sub-networks to extract spatio-temporal features from three types of links. The extracted features are then further processed by fully connected (FC) layers to generate accurate channel-coefficient predictions. Furthermore, we design a channel quality-aware loss function that assigns larger penalties to low-quality links with low average channel power. This biases the training process toward statistically weak links and improves the overall probability of satisfying URLLC reliability requirements.
	
		\item[$\bullet$] 
        We design an MoE network with three lightweight experts for uplink power allocation, each corresponding to a distinct optimization mode, namely EE maximization, SE maximization, or EE--SE trade-off optimization. This objective-aware expert design mitigates the conflicting optimization priorities arising from heterogeneous UE requirements in a single shared model and enables more specialized decision modeling for different UE groups. In addition, an adaptive gating network dynamically fuses the expert outputs to support flexible and robust optimization under heterogeneous service demands.
		
		\item[$\bullet$] 
		Simulation results demonstrate that the proposed CP-Net effectively mitigates channel aging, achieving average NMSE reductions of 51.7\% and 39.5\% over the Kalman and DNN baselines, respectively. Moreover, compared with the SCA and DNN baselines, the proposed framework exhibits superior generalization ability and robustness across different testing scenarios.	

    \end{itemize}
	
	\subsection{Organization}
	The remainder of this paper is organized as follows: In Section \uppercase\expandafter{\romannumeral2}, we present the channel modeling and performance analysis of the aerial–terrestrial CF-mMIMO system under URLLC constraints. Section \uppercase\expandafter{\romannumeral3} introduces the proposed deep learning–based resource optimization framework. In Section \uppercase\expandafter{\romannumeral4}, simulation results are provided to assess the effectiveness of the proposed methods. Finally, Section \uppercase\expandafter{\romannumeral5} concludes the paper. \textit{Notations:} 
	The symbols $(\cdot)^*$, $(\cdot)^H$, and $(\cdot)^T$ denote the complex conjugate, conjugate transpose (Hermitian), and transpose operations, respectively. 
	The notation $||\cdot||$ represents the Euclidean norm. 
	$\mathcal{N}(0,1)$ and $\mathcal{CN}(0,1)$ denote real and complex Gaussian distributions with zero mean and unit variance, respectively. 
	The expectation operator is denoted by $\mathbb{E}[\cdot]$. 
	$\mathbb{R}^{K \times M}$ and $\mathbb{C}^{K \times M}$ represent the sets of real and complex $K \times M$ matrices, respectively.

	\section{SYSTEM MODEL AND PROBLEM FORMULATION}	
	\par
\begin{figure}[t]
		\centering
		\includegraphics[width=\linewidth]{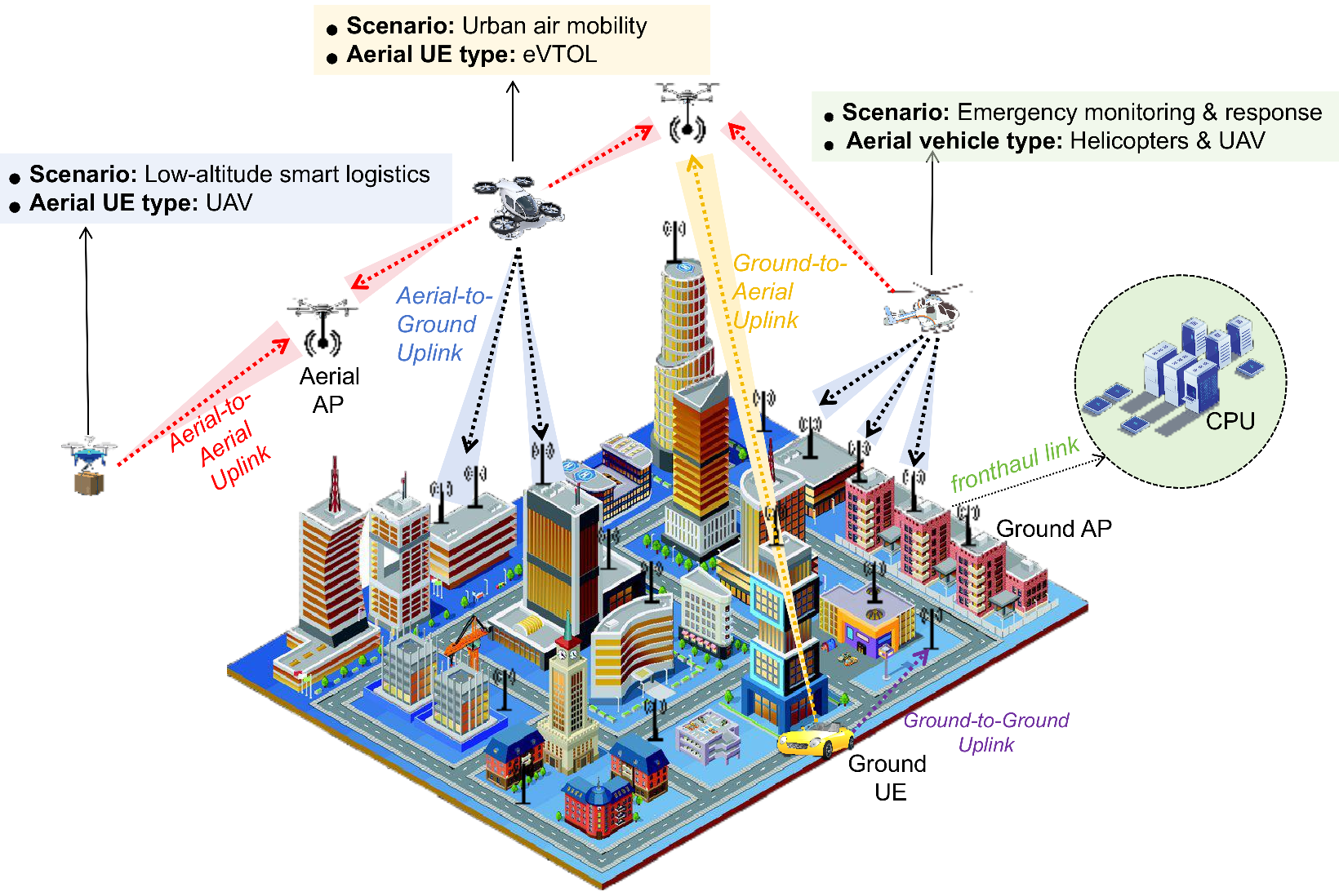}
		\caption{The architecture of a hybrid aerial–terrestrial CF-mMIMO network for supporting URLLC services in UAM scenarios.}
		\label{CF-mMIMO} % 建议添加标签以便引用
	\end{figure}

 We propose a hybrid aerial--terrestrial CF-mMIMO network to support diverse URLLC services in the UAM environment. As shown in Fig.~\ref{CF-mMIMO}, this network consists of $M$ APs, including $m_g$ ground APs and $m_a$ aerial APs. Each AP is equipped with $L$ antennas to serve $K$ single-antenna UEs within the coverage area, including $k_g$ ground UEs and $k_a$ aerial UEs. The received signals at all APs are forwarded to a CPU via fronthaul links for centralized combining and processing. The uplink transmission follows a time-division duplex (TDD) protocol and is modeled as a block-fading channel. We assume that all UEs share the same time--frequency resources and transmit a short packet of \(D\) bits over \(N = B t_{\text{tr}}\) channel uses (CUs), where \(B\) is the system bandwidth and \(t_{\text{tr}}\) denotes the transmission latency \cite{chen2024energy}.
	
	\subsection{Channel Aging Model}
    \par The proposed CF-mMIMO network considers three types of uplink links: aerial-to-aerial (ATA), ground-to-ground (GTG), and air--ground (AG) links, where AG links encompass both aerial-to-ground and ground-to-aerial links \cite{zheng2021uav}. Specifically, Rician fading is adopted for AG links to accurately capture their propagation characteristics, as these links typically exhibit a dominant line-of-sight (LoS) component along with scattered multipath due to relatively open environments with sparse scatterers \cite{9508416, 11045816}. The channel vector between the $k$-th UE and the $m$-th AP at sampling instant $t$, denoted as $\textbf{\textit{h}}^{(\text{h})}_{k,m}[t] \in \mathbb{C}^{L \times 1}$, is expressed as
	
	\begin{equation}
		\small
		\label{eq1}
		\textbf{\textit{h}}^{(\text{h})}_{k,m}[t]=\sqrt{\frac{K^{(\text{h})}_{k,m}}{K^{(\text{h})}_{k,m}+1}} \tilde{\textbf{G}}^{(\text{L})}_{k,m}[t]+\sqrt{\frac{1}{ K^{(\text{h})}_{k,m}+1}}\tilde{\textbf{\textit{G}}}^{(\text{N})}_{k,m}[t],
		\tag{1}
	\end{equation}
	
	\noindent where $ K^{(\text{h})}_{k,m}$ denotes the $K$-factor of AG links, the LoS and non-line-of-sight (NLoS) components are represented by $\tilde{\mathbf{G}}^{(\mathrm{L})}_{k,m}[t] = \textbf{\textit{g}}^{(\mathrm{L})}_{k,m}[t] d_{k,m}[t]^{-A^{(\text{L})}_{h}/2}$ and $\tilde{\mathbf{G}}^{(\mathrm{N})}_{k,m}[t] = \textbf{\textit{g}}^{(\mathrm{N})}_{k,m}[t] d_{k,m}[t]^{-A^{(\text{N})}_{h}/2}$, respectively. Here, \(\textbf{\textit{g}}_{k,m}^{(\mathrm{L})}[t]=e^{j\phi_{k,m}}\mathbf{a}_{k,m}^{(\mathrm{L})}[t]\) denotes the LoS channel vector, where \(\mathbf{a}_{k,m}^{(\mathrm{L})}[t]\in \mathbb{C}^{L \times 1}\) is the normalized LoS spatial signature with \(\|\mathbf{a}_{k,m}^{(\mathrm{L})}[t]\|=1\), and \(\phi_{k,m}\) is a random phase uniformly distributed over \([-\pi,\pi)\), $\textbf{\textit{g}}^{(\mathrm{N})}_{k,m}[t]$ represents the NLoS channel vector with independent and identically distributed (i.i.d.) $\mathcal{CN}(0,1)$, $d_{k,m}[t]$ is the distance between the $k$-th UE and the $m$-th AP, $A^{(\text{L})}_{h}$ and $A^{(\text{N})}_{h}$ are the path loss exponents of the LoS and NLoS channels, respectively. In contrast, ATA links generally experience unobstructed aerial propagation conditions and therefore tend to exhibit a strong LoS component, while scattered multipath may still exist due to environmental reflections. Accordingly, the ATA channel can be modeled as
    \textcolor{black}{
    \begin{equation}
		\small
		\label{eq2}
		\textbf{\textit{h}}^{(\text{a})}_{k,m}[t]=\sqrt{\frac{K^{(\text{a})}_{k,m}}{K^{(\text{a})}_{k,m}+1}} \hat{\textbf{G}}^{(\text{L})}_{k,m}[t]+\sqrt{\frac{1}{K^{(\text{a})}_{k,m}+1}}\hat{\textbf{\textit{G}}}^{(\text{N})}_{k,m}[t],
		\tag{2}
	\end{equation}}
    
    \noindent where $K^{(a)}_{k,m}$ denotes the $K$-factor for ATA links and is set to a relatively large value, $\hat{\mathbf{G}}^{(\mathrm{L})}_{k,m}[t]=\mathbf{g}^{(\mathrm{L})}_{k,m}[t]\, d_{k,m}[t]^{-A^{(\mathrm{L})}_{a}/2}$ and $\hat{\mathbf{G}}^{(\mathrm{N})}_{k,m}[t]=\mathbf{g}^{(\mathrm{N})}_{k,m}[t]\, d_{k,m}[t]^{-A^{(\mathrm{N})}_{a}/2}$ represent the LoS and NLoS components of ATA links, respectively, where $A^{(\mathrm{L})}_{a}$ and $A^{(\mathrm{N})}_{a}$ are the corresponding path-loss exponents.
    
    \par GTG channels are modeled by Rayleigh fading to capture the rich scattering in dense terrestrial environments, where diffraction and reflections from surrounding obstacles result in severe multipath propagation and typically preclude a dominant LoS component \cite{10507759}. Accordingly, the GTG channel vector is given by $\textbf{\textit{h}}^{(\text{g})}_{k,m}[t]=\textbf{\textit{g}}^{(\mathrm{N})}_{k,m}[t]\; d_{k,m}[t]^{-A_{g}/2}$, where $A_{g}$ is the path-loss exponent for GTG links \cite{11045816}. \footnote{To ensure notational consistency in the subsequent derivations, unless otherwise specified, we use $\bm h_{k,m}[t]$ as a unified notation to denote the channel vector from the $k$-th UE to the $m$-th AP at time slot $t$, covering the GTG, ATA, and AG links.}
    
	\par In the proposed aerial-terrestrial network, high-mobility UEs induce significant time-selective fading, causing the channel to vary dynamically over coherence intervals (CIs) \cite{chopra2017performance}. To accurately characterize these temporal variations, the channel aging effect is modeled based on the temporal correlation between the current and outdated channel states \cite{10859261}, as \begin{equation} \small \label{eq3} \textbf{\textit{h}}_{k,m}[t] = \rho_k[t-t_0]\textbf{\textit{h}}_{k,m}[t_0] + \hat{\rho}_k[t-t_0]\textbf{\textit{r}}_{k,m}[t], \tag{3} \end{equation}

	\noindent where $\textbf{\textit{h}}_{k,m}[t_0]$ is the initial channel vector at time slot $t_0$, $\textbf{\textit{r}}_{k,m}[t]\in\mathbb{C}^{L\times1}$ represents the innovation component, which captures the random fluctuations of the channel. The distribution of $\textbf{\textit{r}}_{k,m}[t]$ depends on the link type: $\textbf{\textit{r}}_{k,m}[t]\sim\mathcal{CN}(\mathbf{0},d_{k,m}[t]^{-A_{g}}\mathbf{I}_L)$ for GTG links, $\textbf{\textit{r}}_{k,m}[t]\sim\mathcal{CN}(\mathbf{0},\mathbf R^{(\mathrm{h})}_{k,m}[t])$ for AG links, $\textbf{\textit{r}}_{k,m}[t]\sim\mathcal{CN}(\mathbf{0},\mathbf R^{(\mathrm{a})}_{k,m}[t])$ for ATA links, where $\mathbf R^{(\mathrm{h})}_{k,m}[t]=\mathbb{E}\!\left\{\textbf{\textit{h}}^{(\mathrm{h})}_{k,m}[t]\textbf{\textit{h}}^{(\mathrm{h})H}_{k,m}[t]\right\}$ and $\mathbf R^{(\mathrm{a})}_{k,m}[t]=\mathbb{E}\!\left\{\textbf{\textit{h}}^{(\mathrm{a})}_{k,m}[t]\textbf{\textit{h}}^{(\mathrm{a})H}_{k,m}[t]\right\}$. The complementary coefficient $\hat\rho_k[t-t_0]$ is formulated as
	
	\begin{equation}
		\small
		\label{eq4}
		\hat\rho_k[t-t_0]=\sqrt{1-\rho_k^2[t-t_0]},
		\tag{4}
	\end{equation}
	
	\noindent where $\rho_{k}[t-t_{0}] = J_0(2\pi \mathcal{D}_k \mathcal{F} |t-t_{0}|)$, with $J_0(\cdot)$ denoting the zeroth-order Bessel function of the first kind, $\mathcal{D}_k = \frac{v_k f_c}{v_c}$ representing the maximum Doppler frequency. Here, $v_k$, $f_c$, $\mathcal{F}$, and $v_c =3.0 \times 10^8$~m/s are the UE velocity, carrier frequency, sampling interval, and speed of light, respectively.\footnote{The adopted channel and aging models are used as tractable approximations to capture the heterogeneous link conditions and mobility-induced CSI aging in the considered UAM scenario, while also providing a stable and reproducible modeling basis for the training and evaluation of the proposed learning framework.}
	
	\subsection{Channel Estimation and Multi-User Detection}
	
	\par Owing to the multipath propagation, obstacles, and user mobility, the wireless channel typically experiences fast fading, which necessitates accurate and timely channel estimation \cite{zeng2023achieving}. To address this challenge, we propose an efficient short-packet structure that encapsulates the pilot signal $\bm{\phi}_{k}[t] \in \mathbb{C}^{1 \times \tau_{\text{p}}}$ and the data signal $\bm{x}_{k}[t] \in \mathbb{C}^{1 \times \tau_{\text{s}}}$ within the same packet, assigning blocklengths of $\tau_{\text{p}}$ and $\tau_{\text{s}} = N - \tau_{\text{p}}$, respectively. Here, $N$ denotes the number of channel uses within a packet, and the pilot sequences are transmitted with the same power as the data signals for each UE \cite{10924175}. Furthermore, to avoid the impact of pilot pollution, we assign orthogonal pilot signals to different UEs, i.e., $\bm{\phi}_{k}[t]\bm{\phi}_{k}[t]^H=\tau_{\text{p}}$, and $\bm{\phi}_{k}[t]\bm{\phi}_{k'\neq k}^H[t]=0$. Moreover, the data symbols are mutually independent, such that  $\mathbb{E}\{[\bm{x}_{k}[t]]_{i}[\bm{x}_{k}[t]]_{j}^*\}=0$ for $ i\neq j$. The received signal matrix at the $t$-th instant at the $m$-th AP is expressed by 
	
	\begin{equation}
		\label{eq5}
		\small
		\begin{aligned}[t]
			\mathbf{Y}_{m}[t]=\left[\mathbf{Y}_{\text{p},m}[t],\mathbf{Y}_{\text{s},m}[t]\right] \in \mathbb{C}^{L \times N }, 
		\end{aligned}
		\tag{5}
	\end{equation}
	
	\noindent where $\mathbf{Y}_{\text{p},m}[t]=\sum_{k=1}^K \sqrt{p_k}\,{\textbf{\textit{h}}}_{k,m}[t]\bm{\phi}_{k}[t]+\mathbf{N}_{\text{p},m}[t]$ and $\mathbf{Y}_{\text{s},m}[t]=\sum_{k=1}^K \sqrt{p_k}\,{\textbf{\textit{h}}}_{k,m}[t]\bm{x}_{k}[t]+\mathbf{N}_{\text{s},m}[t]$. Here, $\mathbf{N}_{\text{p},m}[t] \in \mathbb{C}^{L \times \tau_{\text{p}}}$ and $\mathbf{N}_{\text{s},m}[t] \in \mathbb{C}^{L \times \tau_{\text{s}}}$ denote the corresponding noise matrices for pilot and data transmissions, respectively, whose entries are i.i.d.\ $\mathcal{CN}(0,1)$.
	
    \par Subsequently, each AP employs the least squares (LS) method for channel estimation based on the received pilot signals. This approach offers lower computational complexity than alternative techniques such as compressed sensing or minimum mean squared error (MMSE) estimation, which makes it particularly suitable for deployments with a large number of UEs and APs \cite{10706806}. Let $\lambda$ denote the reference instant for channel estimation. The estimated channel from the $k$-th UE to the $m$-th AP at instant $\lambda$ is given by

\begin{equation}
	\label{eq6}
	\small
	\tag{6}
	\begin{aligned}
		\hat{\textbf{\textit{h}}}_{k,m}[\lambda]
		&=\mathbf{Y}_{\text{p},m}[\lambda]\frac{\bm{\phi}^{H}_{k}[\lambda]}{\sqrt{p_{k}}\tau_{\text{p}}} \\
		&=\left(\sum_{i=1}^K \sqrt{p_i}\,{\textbf{\textit{h}}}_{i,m}[\lambda]\bm{\phi}_{i}[\lambda]+\mathbf{N}_{\text{p},m}[\lambda]\right)
		\frac{\bm{\phi}^{H}_{k}[\lambda]}{\sqrt{p_{k}}\tau_{\text{p}}} \\
		&={\textbf{\textit{h}}}_{k,m}[\lambda]+\mathbf{E}_{k,m}[\lambda],
	\end{aligned}
\end{equation}

	\begin{figure*}[!b]
		\begin{center}
			\begin{equation}
				\label{eq9}
				\small
				\begin{aligned}[t]
					\gamma_k[t]=\frac{|\mathbb {E}\{DS_k[t]\}|^2}{\sum_{\substack{i=1}}^{K}\mathbb {E}\{|EL_{k,i}[t]|^2\}+\sum_{\substack{i=1}}^{K}\mathbb {E}\{|AL_{k,i}[t]|^2\}+\mathbb {E}\{|AW_k[t]|^2\}+\sum_{\substack{i=1\\i \neq k}}^{K}\mathbb {E}\{|UI_{k,i}[t]|^2\}}.
				\end{aligned}
				\tag{9} % 手动设置方程的编号为6
			\end{equation}
		\end{center}
	\end{figure*}

	\begin{figure*}[!b]
		\begin{center}
	\begin{equation}
	\label{1SINR}
	\tag{10}
	\small
	\begin{aligned}
		\gamma_k[t]=
		\frac{p_k \rho_k^2[t-\lambda]\left|\sum_{m=1}^{M}\Psi_{k,m}[\lambda]\Theta_{k,m}[\lambda]\right|^2}
		{\sum_{m=1}^{M}\Psi_{k,m}^2[\lambda]\Big(
			\sum_{i=1}^{K}\tau_{\text{p}}^{-1}\rho_i^2[t-\lambda]\Theta_{k,m}[\lambda]
			+\sum_{\substack{i=1\\ i\neq k}}^{K}p_i\rho_i^2[t-\lambda]\Phi_{k,i,m}[\lambda]
			+\sum_{i=1}^{K}p_i\hat{\rho}_i^2[t-\lambda]\mu_{k,i,m}[t]
			+\Theta_{k,m}[\lambda]\Big)}.
	\end{aligned}
	\end{equation}

		\end{center}
	\end{figure*}

	\noindent where $\mathbf{E}_{k,m}[\lambda]=\frac{\mathbf{N}_{\text{p},m}[\lambda]\bm{\phi}^{H}_{k}[\lambda]}{\sqrt{p_{k}}\tau_{\text{p}}}\sim\mathcal{CN}\!\left(\mathbf{0},\frac{1}{p_k\tau_{\mathrm p}}\mathbf{I}_L\right)$ denotes the estimation error. Then, the actual channel at the $\lambda$-th sampling instant can be expressed as $\textbf{\textit{h}}_{k,m}[\lambda]=	\hat{\textbf{\textit{h}}}_{k,m}[\lambda]-\mathbf{E}_{k,m}[\lambda]$.
	
 	\par In the CF-mMIMO architecture, simultaneous transmissions from all UEs over the same time-frequency resources result in significant IUI. Therefore, multi-user detection (MUD) is required to separate the desired signal from IUI \cite{10461045}. In conventional CF-mMIMO systems, the combining schemes such as zero-forcing (ZF), minimum mean square error (MMSE), and large-scale fading decoding (LSFD) may achieve better interference suppression, but they also incur higher computational complexity, require more global channel state information or statistical information, and involve heavier centralized processing \cite{10975288}. Therefore, we employ a two-stage MUD scheme \cite{zeng2023achieving}. In the first stage, each AP receives signals from all UEs and applies the maximum-ratio combining (MRC) to detect the desired signals. The detection vectors are constructed based on the estimated CSI. Specifically, to detect the signal transmitted by the $k$-th UE at the $t$-th instant, the post-processed signal at the $m$-th AP is given by $\hat{\mathbf{y}}_{k,m}[t]=\hat{\textbf{\textit{h}}}_{k,m}[\lambda]^H\mathbf{Y}_{\text{s},m}[t]  \in \mathbb{C}^{1 \times \tau_{\text{s}}}$. In the second stage, we employ a normalized channel-weighting method on the CPU to coherently combine signals from all APs, thereby improving overall signal quality. The combining weight for the \(k\)-th UE at the \(m\)-th AP is defined based on the estimated channel $\hat{\textbf{\textit{h}}}_{k,m}[\lambda]$, i.e., $\Psi_{k,m}[\lambda] = \frac{\mathbb{E}\{\|\hat{\textbf{\textit{h}}}_{k,m}[\lambda]\|^2\}}{\sum_{m'=1}^{M}\mathbb{E}\{\| \hat{\textbf{\textit{h}}}_{k,m'}[\lambda]\|^2\}}$. Accordingly, the combined detection signal for the \(k\)-th UE at the CPU is expressed as
	
	\begin{equation}
		\small
		\label{eq7}
		\begin{aligned}[t]
			\mathbf{y}_{k}[t]&=\sum_{m=1}^{M}\Psi_{k,m}[\lambda]\hat{\mathbf{y}}_{k,m}[t]\\	
			&=\sum_{m=1}^{M}\Psi_{k,m}[\lambda]	\hat{\textbf{\textit{h}}}_{k,m}^H[\lambda]\sum_{i=1}^K \left(\sqrt{p_i}{\textbf{\textit{h}}}_{i,m}[t]\bm{x}_{i}[t]\right)\\
			&+\sum_{m=1}^{M}\Psi_{k,m}[\lambda]	\hat{\textbf{\textit{h}}}_{k,m}^H[\lambda]\mathbf{N}_{\text{s},m}[t].
		\end{aligned}
		\tag{7}
	\end{equation}
	
	\noindent Next, based on (\ref{eq3}) and (\ref{eq6}), $\mathbf{y}_{k}[t]$ can be expanded into the following components:
	
	\begin{equation}
		\label{eq8}
		\small
		\begin{aligned}[t]
			\mathbf{y}_{k}[t]&=\underbrace{\sqrt{p_k}\rho_{k}{[t-\lambda]}\sum_{m=1}^{M}\Psi_{k,m}[\lambda]	\hat{\textbf{\textit{h}}}_{k,m}^H[\lambda]\hat{\textbf{\textit{h}}}_{k,m}[\lambda]}_{DS_k[t]}\bm{x}_{k}[t]\\		
			&-\sum_{\substack{i=1}}^K\underbrace{\sqrt{p_i}\rho_{i}{[t-\lambda]}\sum_{m=1}^{M}\Psi_{k,m}[\lambda]\hat{\textbf{\textit{h}}}_{k,m}^H[\lambda]\mathbf{E}_{i,m}[\lambda]}_{EL_{k,i}[t]}\bm{x}_{i}[t]\\
			&+\sum_{\substack{i=1}}^{K}\underbrace{\sqrt{p_i}\hat\rho_{i}{[t-\lambda]}\sum_{m=1}^{M}\Psi_{k,m}[\lambda]\hat{\textbf{\textit{h}}}_{k,m}^H[\lambda]\textbf{\textit{r}}_{i,m}[t]}_{AL_{k,i}[t]}\bm{x}_{i}[t]\\
			&+\sum_{\substack{i=1\\i \neq k}}^{K}\underbrace{\sqrt{p_{i}}\rho_{i}{[t-\lambda]}\sum_{m=1}^{M} \Psi_{k,m}[\lambda]\hat{\textbf{\textit{h}}}_{k,m}^H[\lambda] \hat{\textbf{\textit{h}}}_{i,m}[\lambda]}_{UI_{k,i}[t]}\bm{x}_{i}[t]\\
			&+\underbrace{\sum_{m=1}^{M} \Psi_{k,m}[\lambda] \hat{\textbf{\textit{h}}}_{k,m}^H[\lambda] \mathbf{N}_{\text{s},m}[t]}_{AW_k[t]},
		\end{aligned}
		\tag{8}
	\end{equation}

	\noindent where $DS_k[t]$ denotes the desired signal of the $k$-th UE, while $UI_{k,i}[t]$, $EL_{k,i}[t]$, $AL_{k,i}[t]$, and $AW_k[t]$ are treated as effective noise terms, corresponding to IUI, estimation error, channel-aging impairment, and AWGN, respectively \cite{chopra2017performance}.

	\par Assuming that the channel statistics are known at the receiver \cite{zeng2023achieving}, the SINR for the $k$-th UE at instant $t$ is formulated in \eqref{eq9}. For simplicity, we define 
	$\Theta_{k,m}[\lambda]=\mathbb{E}\!\left\{\left\|\hat{\textbf{\textit{h}}}_{k,m}[\lambda]\right\|^2\right\}$,
	$\Phi_{k,i,m}[\lambda]=\mathbb{E}\!\left\{\left|\hat{\textbf{\textit{h}}}_{k,m}^H[\lambda]\hat{\textbf{\textit{h}}}_{i,m}[\lambda]\right|^2\right\}$,
	and $\mu_{k,i,m}[t]=\mathbb{E}\!\left\{\left|\hat{\textbf{\textit{h}}}_{k,m}^H[\lambda]\textbf{\textit{r}}_{i,m}[t]\right|^2\right\}$.
	Then, \eqref{eq9} can be rewritten as \eqref{1SINR}.

	\subsection{Performance Analysis}
	
	\par To satisfy the stringent requirements of URLLC, conventional methods such as power control, network densification, and frequent channel estimation have been widely adopted. However, these approaches inevitably incur increased energy consumption and spectrum overhead \cite{ngo2017cell}. In this subsection, we first formalize the latency requirements, then derive a closed-form expression for communication reliability based on FBL theory \cite{zhang2025delay}, and finally analytically evaluate the uplink SE and EE. The latency constraint in URLLC is characterized by the end-to-end latency, denoted by $t_{\mathrm{en}}$, which is defined as the total latency from the transmitter to the receiver:
    \begin{equation}
    \small
    \tag{11}
    t_{\mathrm{en}} = t_{\mathrm{tr}} + t_{\mathrm{pr}} + t_{\mathrm{pc}} + t_{\mathrm{re}} + t_{\mathrm{fh}},
    \end{equation}
    \noindent where $t_{\mathrm{tr}}$, $t_{\mathrm{pr}}$, $t_{\mathrm{pc}}$, $t_{\mathrm{re}}$, and $t_{\mathrm{fh}}$ denote the wireless transmission latency, propagation latency, processing latency, retransmission latency, and fronthaul latency, respectively. In the considered CF-mMIMO system, where APs are densely deployed and communication distances are short, the propagation latency is typically on the order of microseconds ($\mu$s) \cite{10924175}. Since the non-transmission latency components depend strongly on implementation details, they are collectively represented by the overhead term $t_{\mathrm{oh}} = t_{\mathrm{pr}} + t_{\mathrm{pc}} + t_{\mathrm{re}} + t_{\mathrm{fh}}$ in the adopted model. Therefore, this paper focuses on transmission-related optimization. In the simulations, different transmission-latency budgets are considered to evaluate their impact on system performance \cite{zeng2023achieving}.

	\par To ensure both strict low latency and high reliability under URLLC requirements, short-packet transmission schemes are typically employed. The classical Shannon capacity, which assumes asymptotically large blocklengths, becomes inadequate in practical short-packet scenarios \cite{zhang2025delay}. Instead, the error probability (EP) \footnote{In the considered short-packet setting, each packet is transmitted within one time--frequency block of $N$ CUs and decoded as a whole. Therefore, the EP $\varepsilon_k[t]$ in the FBL formulation can be interpreted as the packet error probability for a single transmission attempt.}, derived from FBL theory, is adopted to characterize reliability. EP quantifies the likelihood of decoding failure for fixed-length packets, providing a more accurate and practical performance metric. Based on FBL theory, the achievable data rate (in bits/CU) for the $k$-th UE at the $t$-th instant is given by

	\begin{equation}
		\label{eq12}
		\small
		\begin{aligned}[t]
			\begin{aligned}
				R_k[t]=C(\gamma_{k}[t])-\sqrt{\frac{V(\gamma_{k}[t])}{\tau_{\text{s}}}}Q^{-1}(\epsilon_{k}[t]),
			\end{aligned}
		\end{aligned}
		\tag{12}
	\end{equation}
	
	\noindent where $Q^{-1}(\cdot)$ denotes the inverse $Q$-function. The $Q$-function is defined as $Q(x)=\frac{1}{\sqrt{2\pi}}\int_{x}^{\infty} e^{-\frac{u^{2}}{2}}\,\mathrm{d}u$.
	Moreover, $C(\gamma_k[t])=\log_{2}(1+\gamma_k[t])$ and
	$V(\gamma_k[t])=\frac{\gamma_k[t]^{2}+2\gamma_k[t]}{(1+\gamma_k[t])^{2}}\log_{2}^{2}(e)$
	denote the Shannon capacity and the channel dispersion corresponding to SINR $\gamma_k[t]$, respectively. Consequently, the EP of the $k$-th UE at the $t$-th instant can be calculated as 

	\begin{equation}
	\label{eq13}
	\small
	\begin{aligned}[t]
		\epsilon_k[t]=Q\!\left(\left(C(\gamma_{k}[t])-R_k[t]\right)\sqrt{\frac{\tau_{\text{s}}}{{V(\gamma_{k}[t])}}}\right).
	\end{aligned}
	\tag{13}
	\end{equation}

	\par As mentioned earlier, implementing URLLC incurs considerable spectral and power resource overhead, creating challenges for both EE and SE. In particular, SE is a fundamental metric for characterizing the achievable rate under FBL constraints \cite{pala2023spectral}. Accordingly, the uplink SE of the $k$-th UE at instant $t$ is formally expressed (in bps/Hz) as

	\begin{equation}
	\label{eq14}
	\small
	\begin{aligned}[t]
		\eta_k[t]=R_k[t].
	\end{aligned}
	\tag{14}
	\end{equation}

	\par However, achieving higher data rates generally necessitates increased transmission power, which consequently reduces the system’s EE \cite{li2023joint}. In this paper, the uplink EE of the $k$-th UE at sampling instant $t$ is defined as the amount of information transmitted per unit of energy consumed, measured in bits per joule (bits/J), and is given by
	
	\begin{equation}
		\label{eq15}
		\small
		\begin{aligned}[t]
			\omega_k[t]=\frac{B R_k[t]}{p_k\xi_k+u_{k}},
		\end{aligned}
		\tag{15}
	\end{equation}
	
	\noindent where $\xi_k$ is the reciprocal of the drain efficiency of the $k$-th UE's power amplifier, and $u_{k}$ denotes the static power consumed by the transmitter circuitry of the $k$-th UE \cite{zeng2023achieving}.

	\par As observed from \eqref{eq12}–\eqref{eq15}, the various performance metrics are intricately interdependent, resulting in complex and non-convex relationships. For the target UE, both SE and data rate increase with the SINR when the target EP is fixed. Conversely, if the data rate is fixed, the EP decreases as the SINR increases. Since the SINR depends critically on the underlying channel quality, accurate CSI is essential for reliable performance evaluation and power adaptation. However, channel aging causes CSI mismatch and may severely distort the instantaneous SINR estimation. Therefore, channel prediction is introduced to compensate for outdated CSI and to provide more accurate channel information for subsequent power optimization. However, increasing the transmit power also enlarges the denominator of the EE metric, thereby creating an inherent trade-off between EE and SE. Moreover, higher transmit powers of other UEs aggravate IUI, which further degrades the performance of the target UE. Therefore, since transmit power is tightly coupled with multiple heterogeneous performance metrics, jointly optimizing EE and SE under stringent URLLC constraints is highly challenging.
	
	\subsection{Problem Formulation}

	\par To address the growing diversity of applications in low-altitude aerial-terrestrial networks, this paper focuses on uplink optimization of both EE and SE under URLLC constraints. Three distinct optimization objectives are considered to address different application requirements: maximizing SE, maximizing EE, and balancing the trade-off between EE and SE. Let $\mathcal{U}_{SE}$, $\mathcal{U}_{EE}$, and $\mathcal{U}_{EE-SE}$ denote the sets of UEs corresponding to these objectives, respectively. To account for the inter-dependencies among these performance metrics, we introduce a joint optimization indicator to quantify the overall system performance, defined as

\begin{equation}
\label{eq16}
\small
\delta[t]=\frac{\beta_1 \sum_{i\in\mathcal{U}_{SE}}\tilde{\eta}_i[t]}{|\mathcal{U}_{SE}|}+\frac{\beta_2 \sum_{i\in\mathcal{U}_{EE}}\tilde{\omega}_i[t]}{|\mathcal{U}_{EE}|}+\frac{\beta_3 \sum_{i\in\mathcal{U}_{EE-SE}}\tilde{\zeta}_i[t]}{|\mathcal{U}_{EE-SE}|}.
\tag{16}
\end{equation}

\noindent Here, $\tilde{\eta}_i[t] = \frac{\eta_i[t]}{\eta_{\max}}$ and
$\tilde{\omega}_i[t] = \frac{\omega_i[t]}{\omega_{\max}}$, and $\tilde{\zeta}_i[t]= \frac{\tilde{\eta}_i[t]+\tilde{\omega}_i[t]}{2}$ denote the normalized SE, EE, and EE-SE trade-off of the $i$-th UE, respectively, where $\eta_{\max}$ and $\omega_{\max}$ are predefined reference upper bounds under the considered system model and are used solely for normalization. This normalization is adopted to remove the scale mismatch between SE and EE, thereby keeping the corresponding supervision terms at comparable magnitudes and improving training stability. Moreover, $\beta_1$, $\beta_2$, and $\beta_3$ are fixed weighting coefficients that reflect the relative priorities of the SE, EE, and their trade-off service categories, respectively. Based on this joint indicator, the uplink optimization problem is formulated as follows.
	
\begin{subequations}\label{eq:16}  
	\small
	\begin{align}
		\label{eq17}
		\max_{\mathbf{P}=\{p_1, p_2,\ldots, p_K\}} \quad & \delta[t] \tag{17}\\
		\text{s.t.} \quad 
		& 0 < p_k \leq  p_{\max}, \quad \forall k \in \{1,2,\ldots,K\}, \tag{17a}\\
		& \epsilon_k[t]\leq \epsilon_b, \quad \forall k \in \{1,2,\ldots,K\}, \tag{17b}\\
		& t_{\mathrm{en}} \leq t_b, \tag{17c}\\
		& 1 \leq \tau_{\text{s}} \leq N-K. \tag{17d}\label{17d}
	\end{align}
\end{subequations}

	\noindent where $p_{\max}$ denotes the maximum transmit power, $\epsilon_b$ and $t_b$ represent the URLLC boundary of EP and latency, respectively, and \eqref{17d} is imposed to reserve sufficient pilot resources for assigning orthogonal pilot sequences to the $K$ UEs, thereby mitigating pilot pollution.

	\section{Performance optimization by Mixture of Experts Network}
	
	\begin{figure*}[t]
	\centering
	\includegraphics[width=\linewidth]{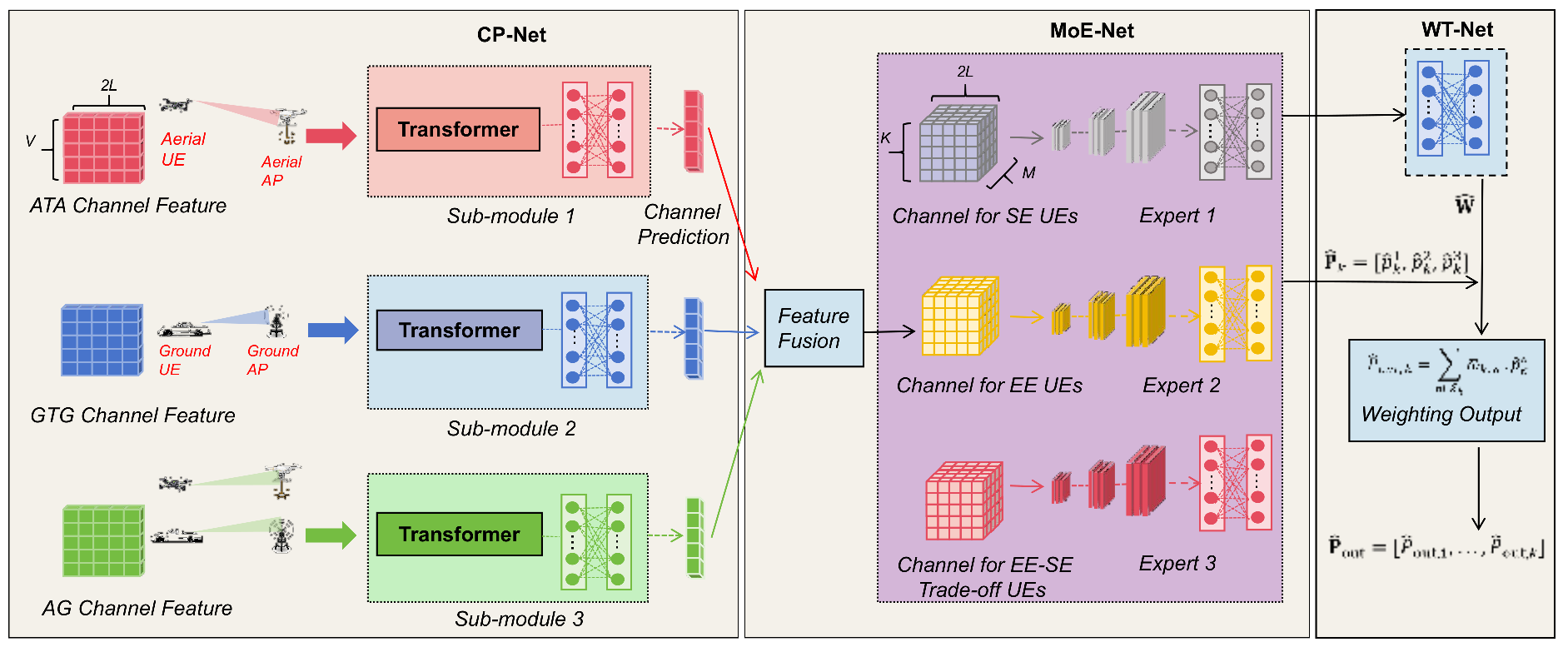}
	\caption{The entire process of the proposed joint networks for channel prediction with CP-Net and power allocation via MoE-Net.}
	\label{fig:2} % 建议添加标签以便引用
	\end{figure*}
	
    \par To address diverse applications and heterogeneous task requirements, it is essential to develop approaches that can jointly optimize multiple objectives. However, existing methods are typically designed for single-objective optimization and rarely address such hybrid optimization problems. Moreover, conventional iterative algorithms often suffer from high computational complexity and slow convergence, which significantly limit their applicability under stringent URLLC constraints. To overcome these challenges, we propose a deep learning-based framework, named CF-MoENet, for solving the uplink power-allocation problem in \eqref{eq17}. As shown in Fig.~\ref{fig:2}, CF-MoENet is a unified framework consisting of three modules: a channel prediction network (CP-Net), a mixture-of-experts power-allocation network (MoE-Net), and a weighting network (WT-Net). Specifically, CP-Net is first employed to predict time-varying channels and mitigate channel-aging effects. Based on the predicted CSI, MoE-Net generates the power-allocation scheme, while WT-Net adaptively determines the fusion weights of different experts. Compared with conventional approaches, the proposed CF-MoENet achieves superior performance and enhanced adaptability through objective-aware expert specialization and adaptive expert fusion. The architecture and key parameters of the proposed networks are summarized in Table~\ref{tab:combined_params}.

\begin{table}[t]
	\centering
	\caption{Architecture and Parameters of the Proposed Networks}
	\label{tab:combined_params}
	\renewcommand{\arraystretch}{1.25}
	\setlength{\tabcolsep}{5pt}
	\footnotesize
	\begin{tabular}{l|c|c}
		\toprule
		\textbf{Network} & \textbf{Description} & \textbf{Dimensions / Parameters} \\
		\hline
		\multirow{9}{*}{CP-Net}
		& Transformer: Number of heads & 4 \\
		& Transformer dimension & $2L$ \\
		& FFN dimension & $4\times 2L$ \\
		\cline{2-3}
		& Input $\mathbf{H}_{k,m}[\lambda]$ & $\mathbb{R}^{V \times 2L}$ \\
		& Transformer Encoder & $V \times 2L$ \\
		& Flatten layer & $(V \times 2L) \rightarrow 2LV$ \\
		& FC Layer 1-1 & $2LV \rightarrow LV$ \\
		& FC Layer 1-2 & $LV \rightarrow 2L$ \\
		& Output $\mathbf{H}_{k,m}^{(\text{p})}[t]$ & $\mathbb{R}^{2L \times 1}$ \\
		\midrule
		\multirow{9}{*}{MoE-Net}
		& Conv Layers: Kernel Size & $3 \times 3$ \\
		& Conv Layer 1 (Filters) & $2M$ \\
		& Conv Layer 2 (Filters) & $4M$ \\
		& Conv Layer 3 (Filters) & $8M$ \\
		\cline{2-3}
		& Input $\mathbf{G}[t]$ & $\mathbb{R}^{K \times M \times 2L}$ \\
		& Expert Network (CNN) & $8M \times 2L$ \\
		& Adaptive Pooling & $(8M \times 2L) \rightarrow 2L$ \\
		& FC Layer 2-1 & $2L \rightarrow L$ \\
		& FC Layer 2-2 & $L \rightarrow 1$ \\
		& Output $\hat{\mathbf{P}}_n$ & $\mathbb{R}^{K \times 1}$ \\
		\midrule
		\multirow{4}{*}{WT-Net}
		& Input $\hat{\mathbf{G}}[t]$ & $\mathbb{R}^{K \times (2ML + 3)}$ \\
		& FC Layer 3-1 & $(2ML + 3) \rightarrow M$ \\
		& FC Layer 3-2 & $M \rightarrow 3$ \\
		& Output $\hat{\mathbf{W}}$ & $\mathbb{R}^{K \times 3}$ \\
		\bottomrule
	\end{tabular}
\end{table}

	\subsection{Deep Learning for Channel Prediction}
	
	\par Imperfect channel estimation can significantly degrade system performance, particularly for UEs with limited transmit power or in severe fading conditions. In such scenarios, conventional power control strategies typically provide only marginal performance gains. To address this challenge, we propose a CP-Net that learns the mapping from past estimated channel states to future channel states affected by aging. CP-Net employs a Transformer-based encoder \cite{han2022survey}, where the multi-head self-attention module extracts complementary temporal features from different representation subspaces and better captures heterogeneous correlation patterns across link types and aging conditions, followed by an FC decoder to reconstruct channel estimates. This architecture effectively mitigates channel aging, enhances estimation accuracy, and improves the performance of subsequent power allocation strategies.

	\par Considering that different types of channels exhibit distinct fading characteristics, we design three sub-modules for channel prediction. This architecture is flexible and scalable, allowing for the addition of new modules to support new channels or objectives. Specifically, Module 1 handles ATA channels, Module 2 targets GTG channels, and Module 3 is dedicated to AG channels. To construct the input features for CP-Net, we first decompose the estimated channel matrix into its real and imaginary components and then concatenate them along the feature dimension. Let $\mathcal{U}_a$, $\mathcal{U}_g$, $\mathcal{A}_a$, and $\mathcal{A}_g$ denote the sets of aerial UEs, ground UEs, aerial APs, and ground APs, respectively. Let $\lambda$ denote the estimation instant. We define $\mathcal{H}_{a2a} = \{\mathbf{H}_{k,m}[\lambda] \mid k \in \mathcal{U}_a, m \in \mathcal{A}_a\}$, $\mathcal{H}_{g2g} = \{\mathbf{H}_{k,m}[\lambda] \mid k \in \mathcal{U}_g, m \in \mathcal{A}_g\}$, and $\mathcal{H}_{hybrid} = \{\mathbf{H}_{k,m}[\lambda] \mid (k \in \mathcal{U}_a, m \in \mathcal{A}_g)\ \text{or}\ (k \in \mathcal{U}_g, m \in \mathcal{A}_a)\}$. These three channel sets are fed into Modules 1, 2, and 3, respectively. Here,
	$\mathbf{H}_{k,m}[\lambda] \in \mathbb{R}^{V \times 2L}$ represents the time-sequential feature between the $k$-th UE and the $m$-th AP over a sequence length of $V$
	and is defined as

	\begin{equation}
	\label{18}
	\small
	\begin{aligned}
		\mathbf{H}_{k,m}[\lambda] = [\mathrm{op}(\hat{\textbf{\textit{h}}}_{k,m}[\lambda-V+1]), \dots, \mathrm{op}(\hat{\textbf{\textit{h}}}_{k,m}[\lambda])]^T,
	\end{aligned}
	\tag{18}
 	\end{equation}

	\noindent where op$(\cdot)$ is a mapping operation that transforms an $L$-dimensional complex-valued CSI vector into a $2L$-dimensional real-valued representation, which is expressed as
	
	\begin{equation}
	\label{19}
	\mathrm{op}(\hat{\textbf{\textit{h}}}_{k,m}[n]) =
	\begin{bmatrix}
		\Re\{\hat{\textbf{\textit{h}}}_{k,m}[n]\} \\
		\Im\{\hat{\textbf{\textit{h}}}_{k,m}[n]\}
	\end{bmatrix}
	\in \mathbb{R}^{2L \times 1}.
		\tag{19}
	\end{equation}
	
	\par Furthermore, CP-Net is trained in a supervised manner. Specifically, the estimated CSI at the actual sampling instant $t=\lambda + t_a$, denoted by \(\hat{\textbf{\textit{h}}}_{k,m}[t]\), is used as the supervisory label, where \(t_a\) denotes the aging interval. These labels explicitly guide the learning process, enabling the network to predict future channel states accurately. The CP-Net architecture comprises a single-layer transformer encoder with four attention heads and a feedforward network (FFN), followed by a two-layer fully connected network that maps the flattened transformer features to the predicted channel. The corresponding mapping function is denoted as $f_{cp}: \mathbf{H}_{k,m}[\lambda] \mapsto \mathbf{H}_{k,m}^{(\text{p})}[t]$, where  $\mathbf{H}_{k,m}^{(\text{p})}[t]=\mathrm{op}(\hat{\textbf{\textit{h}}}^{(p)}_{k,m}[t])$ is the predicted channel coefficient between the $k$-th and the $m$-th AP at the instant $t$.

	\subsection{Deep Learning for Power Allocation}
	
	\par Accurate channel prediction provided by CP-Net enables more effective resource allocation. Nevertheless, in the considered scenario, different UEs have interdependent optimization objectives, which inherently limit the robustness and scalability of single-model approaches. To overcome this limitation, we propose a MoE-Net \cite{10937907} to generate the power allocation strategy in optimization problem (\ref{eq17}). The proposed MoE-Net consists of three expert networks whose outputs are adaptively combined through a gating network. As illustrated in Fig.~2, three expert networks are constructed for SE-oriented, EE-oriented, and SE--EE trade-off-oriented UEs. Let $\mathcal{M}_e=\{1,2,3\}$ denote the expert index set, where Experts 1, 2, and 3 correspond to these three objectives, respectively. Then, for each UE $k$, the target expert index is denoted by $j_k\in\mathcal{M}_e$, with $j_k=1$ for $k\in\mathcal{U}_{SE}$, $j_k=2$ for $k\in\mathcal{U}_{EE}$, and $j_k=3$ for $k\in\mathcal{U}_{EE\text{-}SE}$.

    The experts share the same input tensor $\mathbf{G}[t]\in\mathbb{R}^{K\times M\times 2L}$, obtained by stacking group-wise CSI features along the UE dimension as
	$\mathbf{G}[t]=\big[\mathbf{G}_{\mathrm{SE}}[t],\,\mathbf{G}_{\mathrm{EE}}[t],\,\mathbf{G}_{\mathrm{EE-SE}}[t]\big]$,
	where $\mathbf{G}_{\mathrm{SE}}[t]$, $\mathbf{G}_{\mathrm{EE}}[t]$, and $\mathbf{G}_{\mathrm{EE-SE}}[t]$ collect $\mathbf{H}^{(\mathrm{p})}_{k,m}[t]$ for $k\in\mathcal{U}_{SE}$, $k\in\mathcal{U}_{EE}$, and $k\in\mathcal{U}_{EE\text{-}SE}$, respectively, over all $m\in\mathcal{A}_a\cup\mathcal{A}_g$. Then, to extract spatial features from the input channels, we employ a two-dimensional convolutional neural network (CNN) as the encoder. The encoder consists of three convolutional layers, followed by two FC layers with Sigmoid activations, which map the extracted high-dimensional features to the corresponding power allocation scheme. For the $k$-th UE, the outputs are represented as $\hat{\mathbf{P}}_{k} = [\hat{p}^1_{k}, \hat{p}^2_{k}, \hat{p}^3_{k}]$, where $\hat{p}^n_{k}$ denotes the output of the $n$-th expert.

	\par Moreover, we propose a weighting network (WT-Net) to dynamically assign expert weights for each UE. Specifically, the predicted channel coefficients are first flattened into a one-dimensional vector and concatenated with the predicted power allocation outputs from the three expert models \cite{10937907}. The resulting feature vectors from all UEs are then stacked to form a two-dimensional input matrix $\hat{\mathbf{G}}[t] \in \mathbb{R}^{K \times (2ML + 3)}$, which is subsequently fed into a shared WT-Net consisting of two FC layers followed by a sigmoid activation layer. The output of WT-Net is an expert weighting matrix, denoted by $\hat{\mathbf{W}} = [\hat{\mathbf{W}}_1, \hat{\mathbf{W}}_2, \hat{\mathbf{W}}_3] \in \mathbb{R}^{K \times 3}$, where each column corresponds to the expert weights assigned to all UEs. Furthermore, to enhance expert specialization, for each UE, only the experts with weights exceeding a predefined threshold are selected, while the remaining experts are masked out. The weights corresponding to the selected experts are then re-normalized to ensure that they sum to one, i.e.,
	$\bar w_{k,n}=\frac{ [\hat{\mathbf{W}}_{n}]_k}{\sum_{j\in\mathcal{S}_k} [\hat{\mathbf{W}}_{j}]_k}$ ($n \in \mathcal{S}_k$), where $\mathcal{S}_k$ denotes the index set of the selected experts for the $k$-th UE satisfying $[\hat{\mathbf{W}}_{n}]_k > \tau_w$. If $\mathcal{S}_k=\varnothing$, the expert with the largest weight is retained. The gating threshold $\tau_w$ mainly controls the sparsity of final expert fusion: a larger threshold enforces stronger pruning but may incur some performance loss, whereas a smaller threshold preserves more auxiliary expert information at the expense of weaker sparsity. The overall power allocation is computed as a weighted combination of the expert outputs. For the $k$-th UE, the final predicted power allocation is given by
	$\hat{{p}}_{\text{out},k} = \sum_{n \in \mathcal{S}_k} \bar w_{k,n } \cdot \hat{p}^n_k
	$. The mapping function of the overall power allocation framework is defined as $f_{\mathrm{MoE}}:\mathbf{G}[t]\mapsto \hat{\mathbf{P}}_{\mathrm{out}}$, where $\hat{\mathbf{P}}_{\mathrm{out}}=[\hat{p}_{\mathrm{out},1},\dots,\hat{p}_{\mathrm{out},K}]$ denotes the complete power-allocation vector.

\subsection{Details of Networks}

\par During DNN training, the loss function measures the discrepancy between the predicted outputs and the reference targets, which guides the gradient-based optimization of the network parameters. This enables the network to iteratively adjust its weights to reduce prediction errors and approximate the optimal mapping \cite{10924175}. To effectively learn both channel prediction and power allocation, we divide the training process into two stages, as follows:

\textit{Stage 1: Supervised Training for CP-Net:} In this stage, CP-Net is trained in a supervised manner to learn the mapping from the estimated channel states to the aged channel states. To enhance the prediction accuracy for weak links, we adopt a channel quality-aware weighted MSE loss, where the channel quality is characterized by the average channel power $\Theta_{k,m}[t]$. Accordingly, the weighted CSI prediction loss at instant $t$ is given by
\begin{equation}
	\small
	\mathcal{L}_{\text{CSI}} = \frac{1}{KM} \sum_{k=1}^{K} \sum_{m=1}^{M} \hat\omega_{k,m}[t]
	\left\| \hat{\textbf{\textit{h}}}^{(\text{p})}_{k,m}[t] - \hat{\textbf{\textit{h}}}_{k,m}[t] \right\|^2.
	\tag{20}
\end{equation}
\noindent where $\hat\omega_{k,m}[t] = \frac{1}{\Theta_{k,m}[t]+\hat\epsilon}$, and $\hat\epsilon>0$ is a small constant introduced for numerical stability. In this way, links with lower average channel power are assigned larger loss weights, thereby enabling CP-Net to place greater emphasis on weak links during training.

\textit{Stage 2: Label-Free Model-Driven Training for MoE-Net and WT-Net:} In this stage, the MoE-Net and WT-Net are trained in a label-free, model-driven manner to optimize the uplink power allocation strategy across multiple heterogeneous objectives. Specifically, the network parameters are updated by directly optimizing an analytically constructed loss function, without relying on ground-truth optimal power-allocation labels. The loss function integrates three components: objective maximization, expert specialization to encourage each expert to focus on a specific optimization goal, and URLLC constraint satisfaction to guarantee latency and reliability requirements. Since CP-Net compensates for channel-aging errors, the predicted CSI is used as a surrogate for the aged CSI during this optimization stage. Accordingly, we construct a prediction-based SINR metric for training. Let $O_{k,m}[t] \triangleq \sum_{\substack{i=1 \\ i \neq k}}^{K} \hat p_{\text{out},i}\,\Phi_{k,i,m}[t]$ \cite{zeng2023achieving}. Then, the SINR of the $k$-th UE at instant $t$ based on the predicted CSI is defined as

\begin{equation}
	\label{2SINR}
	\small
	\begin{aligned}
	\gamma_k[t]&=\frac{|\mathbb {E}\{DS_k[t]\}|^2}{\sum\limits_{i=1}^K \mathbb {E}\{|EL_{k,i}[t]|^2\} + \mathbb {E}\{|AW_k[t]|^2\} + \sum\limits_{\substack{i=1\\i \neq k}}^{K} \mathbb {E}\{|UI_{k,i}[t]|^2\}}\\
	&= \frac{\hat p_{\text{out},k} \left|\sum_{m=1}^{M}\Psi_{k,m}[t]\Theta_{k,m}[t]\right|^2}
{\sum_{m=1}^{M}\Psi_{k,m}^2[t]\Big(
	(K\tau_{\text{p}}^{-1}+1)\Theta_{k,m}[t]+O_{k,m}[t]\Big)}.
	\end{aligned}
	\tag{21}
\end{equation}

\noindent Here, $\Psi_{k,m}[t]$, $\Theta_{k,m}[t]$, and $\Phi_{k,i,m}[t]$ are computed from the predicted CSI at instant $t$. Based on the power allocation output by CF-MoENet, the optimized objective value $\delta[t]$ is computed according to (\ref{eq16}). The corresponding objective loss is defined as

\begin{equation}
	\mathcal{L}_{\text{obj}} = (1- \delta[t]).
	\tag{22}
\end{equation}

\par To encourage objective-aware expert specialization, we further introduce a margin-based gating loss imposed directly on the routing weights generated by WT-Net. Specifically, for each UE, this loss penalizes the case where the routing weight of the target expert is not sufficiently larger than those of the non-target experts, thereby enlarging their separation and promoting clearer expert specialization. The corresponding loss is defined as
\begin{equation}
\label{eq23}
\small
	\mathcal{L}_{\text{contrastive}} = \frac{1}{K} \sum_{k=1}^{K} \sum_{j \ne j_k} \max\Big(0, f_{m} - \big([\hat{\mathbf{W}}_{j_k}]_{k} - [\hat{\mathbf{W}}_{j}]_{k}\big)\Big),
\tag{23}
\end{equation}
where $f_m=0.5$ denotes the margin parameter. When $[\hat{\mathbf{W}}_{j_k}]_{k} - [\hat{\mathbf{W}}_{j}]_{k}\ge f_m$, no penalty is incurred, otherwise, the loss remains positive and drives the target expert to receive a larger routing weight than the non-target experts. Furthermore, it is worth noting that while optimizing the uplink performance, the URLLC constraints must be strictly enforced. The corresponding URLLC loss is expressed as
\begin{equation}
	\small
	\label{eq24}
		\mathcal{L}_{\text{URLLC}}
	= \sum_{k=1}^{K}\max(0, \epsilon_k[t]-\epsilon_b).
	\tag{24}
\end{equation}

\par Therefore, the overall multi-task loss function in Stage~2 is formulated as
\begin{equation}\tag{25}\label{eq25}
	\mathcal{L}_{\text{total}}
	= c_1\mathcal{L}_{\text{obj}} + c_2 \mathcal{L}_{\text{contrastive}} + c_3\mathcal{L}_{\text{URLLC}} .
\end{equation}
\noindent where $c_1=c_2=c_3=1/3$ are set to balance the three loss components.

\section{Numerical Results}

In this section, we first provide the experimental design details, including dataset generation, network training, and parameter configurations. Then, we present numerical results to evaluate the performance of the proposed schemes. To construct the experimental dataset, we simulate the proposed hybrid aerial-terrestrial CF-mMIMO system within a $200\, \text{m} \times 200\,\text{m}$ area, where ground UEs and APs are randomly and uniformly distributed. Unless
otherwise specified, the aerial UEs and APs are positioned at random altitudes between $100$~m and $200$~m. To model UE mobility, the velocities of aerial and ground UEs are set to $20$~m/s and $10$~m/s, respectively. The sampling interval $\mathcal{F}$ is 0.1 ms, the aging interval $t_a$ is 8, and gating
threshold $\tau_w$ is 0.1. Under these configurations, we generate the location coordinates of each UE and AP, and compute the corresponding channel coefficients as mentioned. These coefficients are then used as input samples for CP-Net.
\begin{table}[t]
	\centering
	\caption{Simulation Parameters}
	\label{tab2}
	\renewcommand{\arraystretch}{1.1} % 调整行高
	\setlength{\tabcolsep}{8pt}        % 调整列间距
	\begin{tabular}{l|c}                % 添加竖线
		\toprule
		\textbf{Parameters} & \textbf{Value} \\
		\midrule
		Carrier frequency ($f_c$) & 1.9~GHz \\
		$K$-factor of ATA, AG channels ($K^{(\text{a})}_{k,m}$, $K^{(\text{h})}_{k,m}$) & 15, 10 \\
		System bandwidth ($B$) & 10~MHz \\
		Length of data ($D$) & 128~bits \\
		Number of ground APs, aerial APs ($m_g$, $m_a$) & 64, 8 \\
		Number of antennas per AP ($L$) & 4 \\
		Blocklength of pilot ($\tau_{\text{p}}$) & 1500 CU \\
		Latency boundary, Transmission latency ($t_b$, $t_{\mathrm{tr}}$) & 0.5~ms, 0.4 ms\\
        EP boundary ($\epsilon_b$) & $10^{-5}$ \\
        
		Number ratio of aerial and ground UEs & 1:2 \\
		Path loss of LoS components ($A^{(\text{L})}_{h}$, $A^{(\text{L})}_{a}$)  & 2.1, 2.0 \\
        Path loss of NLoS components ($A^{(\text{N})}_{h}$,$A^{(\text{N})}_{a}$, $A^{(\text{N})}_{g}$)
        & 2.5, 2.3, 2.7\\
		Number ratio of $U_{\text{SE}}$, $U_{\text{EE}}$, and $U_{\text{EE-SE}}$ & 1:1:1 \\
		Sequence length of channel ($V$) & 20 \\
		\bottomrule
	\end{tabular}
\end{table}

\begin{figure*}[t]
    \centering
    \subfloat[ATA channels]{
        \includegraphics[width=0.31\linewidth]{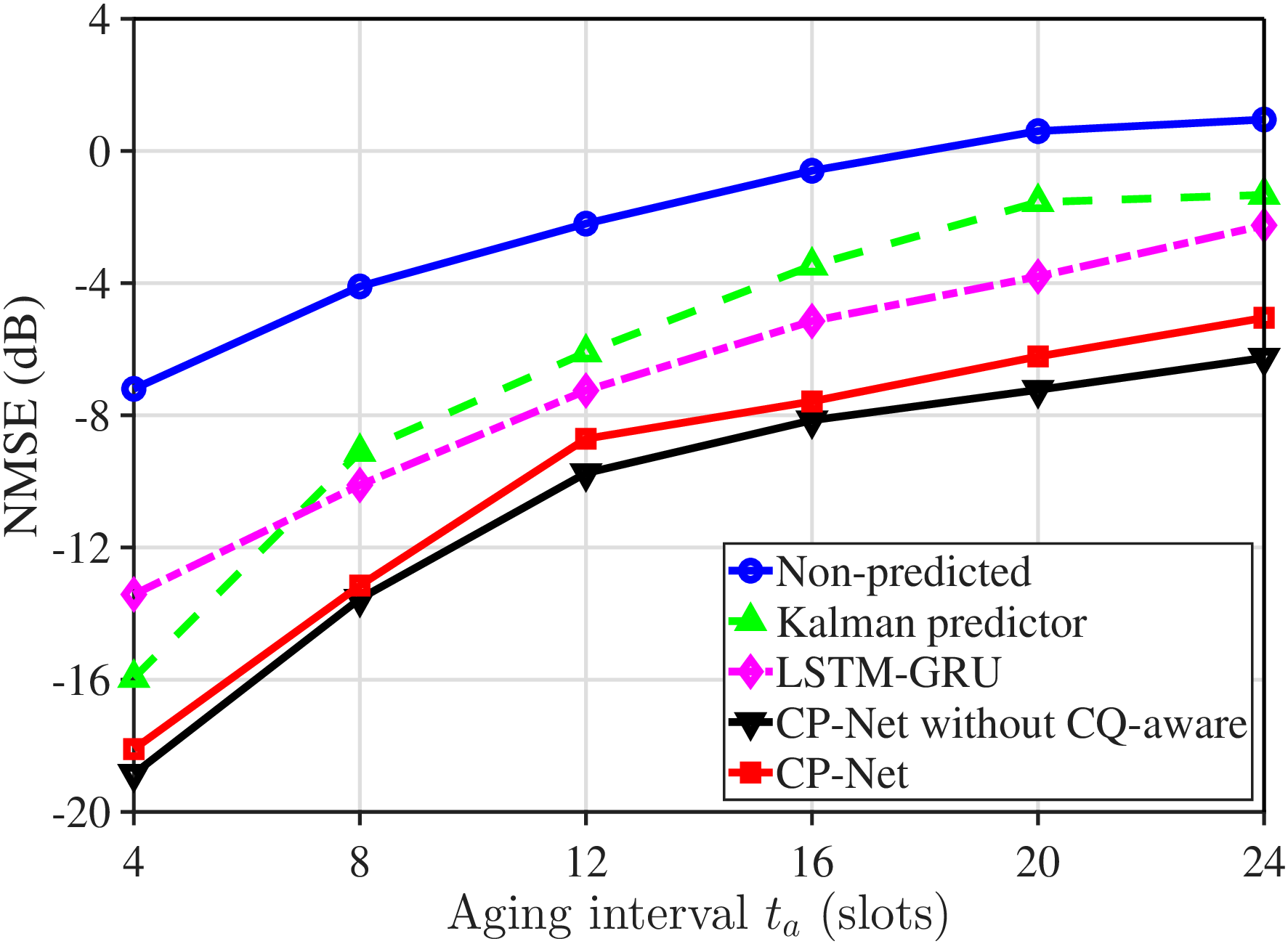}
        \label{fig3a}
    }
    \hfill
    \subfloat[AG channels]{
        \includegraphics[width=0.31\linewidth]{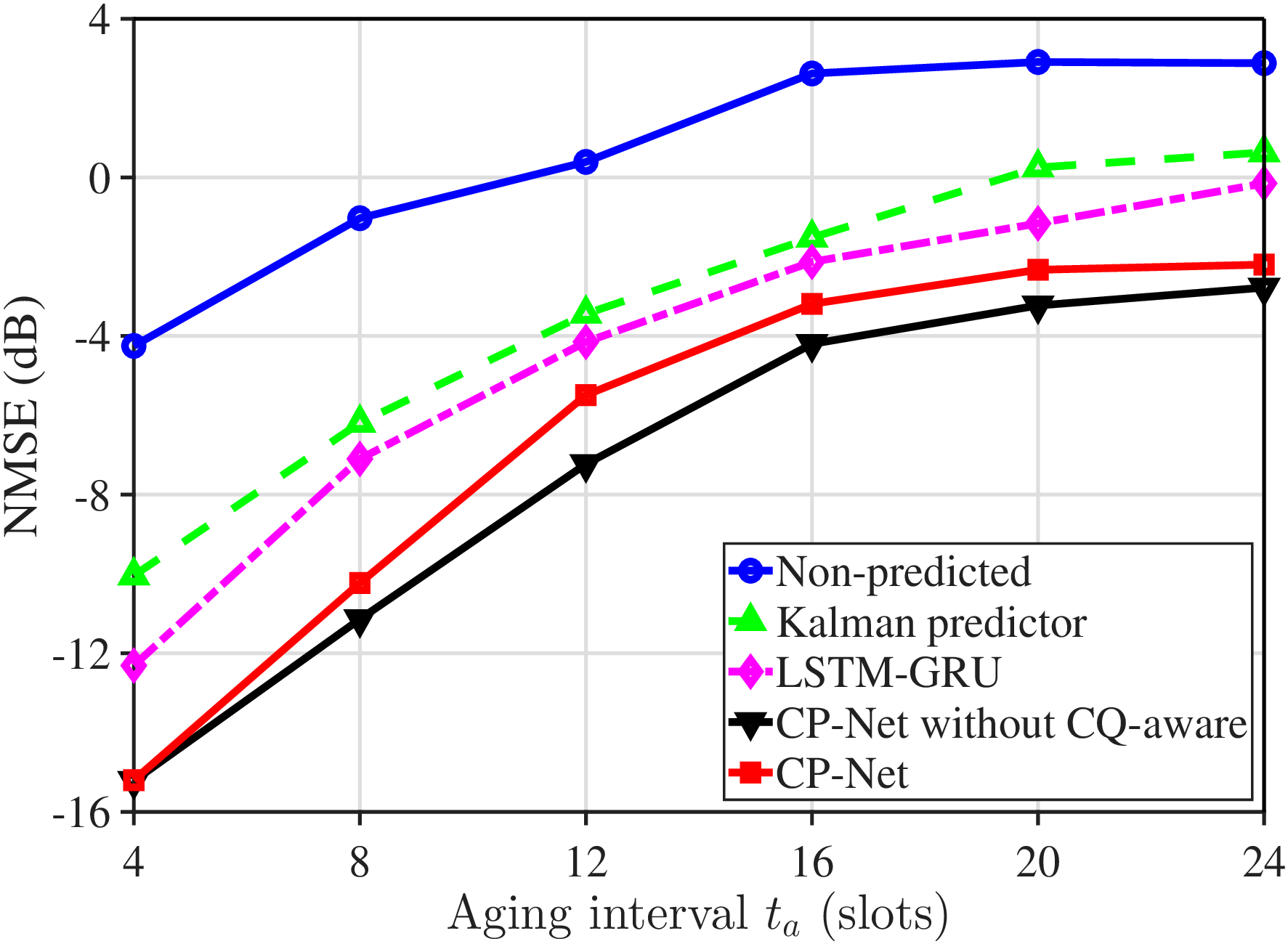}
        \label{fig3b}
    }
    \hfill
    \subfloat[GTG channels]{
        \includegraphics[width=0.31\linewidth]{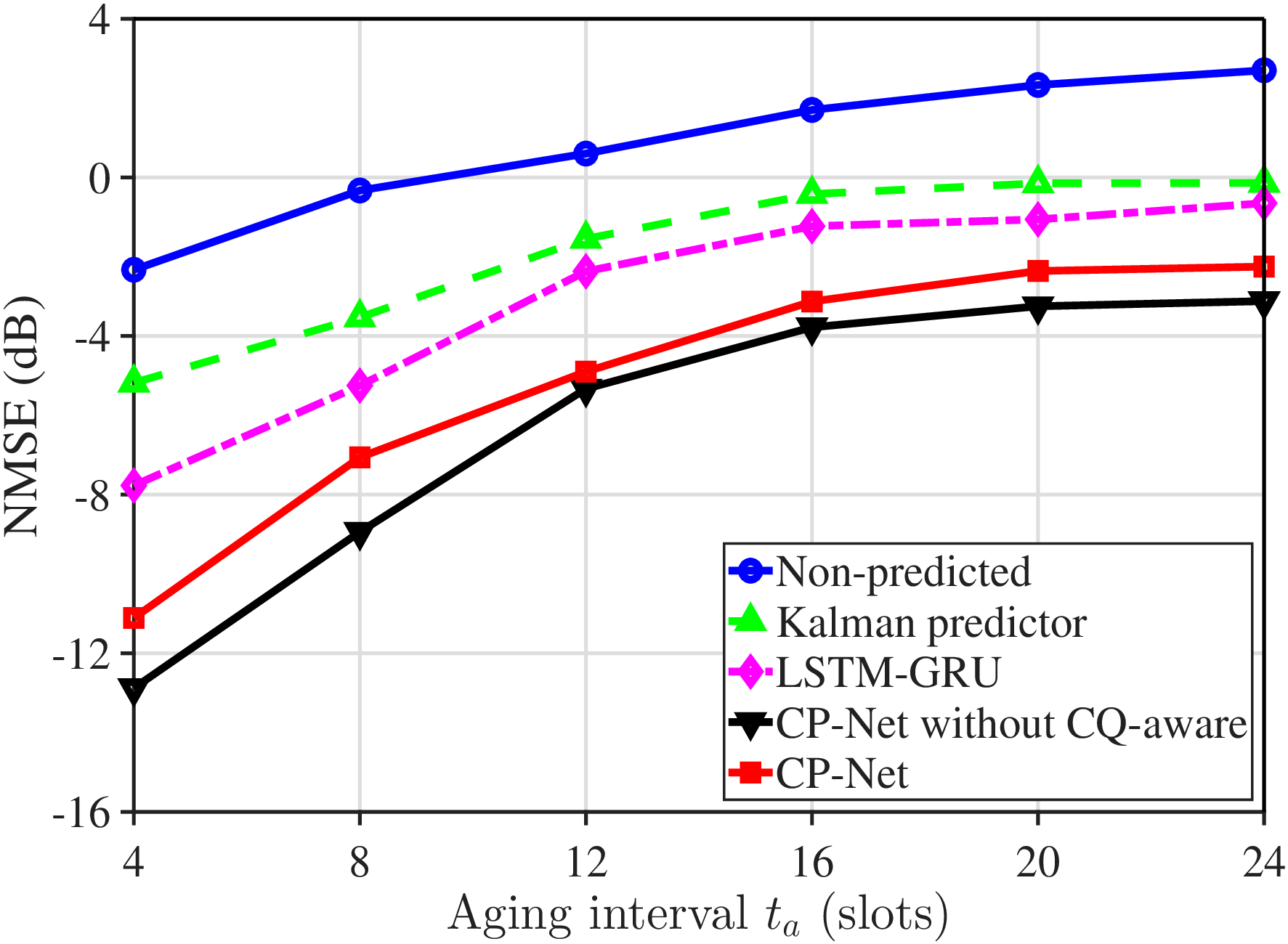}
        \label{fig3c}
    }
    \caption{NMSE performance comparison versus aging interval for different methods in ATA, AG, and GTG channels.}
    \label{fig3}
\end{figure*}

\subsection{Experimental Setup}

\par A total of 20,000 samples are generated and divided into training and testing sets with an 8:2 ratio. The sampling interval is 0.1 ms. The CP-Net and MoE-Net models are trained for 200 epochs with a batch size of 32 and an initial learning rate of 0.01. The optimizing weights are set as $\beta_1=\beta_2=\beta_3=1/3$. Detailed simulation parameters are summarized in Table~\ref{tab2} \cite{van2020joint}. Furthermore, in the FBL regime, the achievable rate $R_k[t]$ and error probability $\epsilon_k[t]$ are inherently coupled. For consistent and fair evaluation, we fix $\epsilon_k[t]=\epsilon_b$ when computing $R_k[t]$ and fix $R_k[t]=R_b$ when calculating $\epsilon_k[t]$. This provides a rigorous basis for jointly optimizing rate and reliability. For the network settings, the experiments were conducted on a workstation with an Intel(R) Core i7 2.10~GHz CPU, 32~GB RAM, and an NVIDIA GeForce RTX 4060 GPU under Windows 11 (64-bit). We design and evaluate a set of baseline methods. For channel prediction, we include an ablation variant of CP-Net that removes the channel quality–weighted loss to quantify its contribution, and we further consider the following predictors:
\begin{itemize}
\item[$\bullet$] \textit{Nonlinear Kalman predictor:}
We adopt a classical nonlinear Kalman predictor \cite{7952818} as a baseline, with parameters updated online during operation.
\item[$\bullet$] \textit{LSTM--GRU predictor:}
Following the architecture in \cite{10089512}, we implement a lightweight cascaded LSTM--GRU model as a deep-learning baseline. By serially combining LSTM and GRU units, it alleviates vanishing-gradient issues and improves convergence.
\end{itemize}

\par For power allocation, we consider the following ablation variants: (i) MoE-Net operating on the estimated channel without channel prediction, (ii) a variant that replaces WT-Net with simple average fusion for decision aggregation, and (iii) a single-expert CNN model. In addition, we include the following baseline methods for comparison:
\begin{itemize}
\item[$\bullet$] \textit{SCA algorithm:}
We employ SCA method \cite{10366311} as an iterative benchmark, where problem \eqref{eq17} is successively approximated by convex subproblems. At each iteration, non-convex terms are locally convexified around the current point, and the resulting convex program is solved using CVX.
\item[$\bullet$] \textit{Lightweight MLP network:}
Following \cite{10045791}, we implement a lightweight MLP for power allocation, consisting of a two-layer fully connected decoder.\footnote{All the deep-learning baselines are implemented as independently trained models, whose architectures and training hyperparameters are tuned for the considered task under the same dataset split, optimization settings, and stopping criterion as the proposed framework.}
\end{itemize}

\par For oracle estimated CSI, the SINR accounting for channel aging effects, as given in Eq.~\eqref{1SINR}, is used to compute the performance metrics. For predicted  CSI, the SINR is computed based on both the predicted and aged channels using the formulation in Eq.~\eqref{2SINR}, enabling a comprehensive assessment of channel prediction accuracy and its impact on system performance.

\subsection{Validation of Channel Prediction}

\par Fig.~\ref{fig3a}-Fig.~\ref{fig3c} compare the normalized mean squared error (NMSE) performance of different prediction methods under various channel types and aging intervals, where the NMSE is computed with respect to the future estimated CSI used as the prediction target. As the aging interval increases, the NMSE of all methods generally rises and gradually levels off. This is because the channel temporal correlation keeps decreasing, thereby reducing the amount of useful information available for prediction. Compared with the baseline schemes, the proposed CP-Net achieves superior prediction accuracy. For example, when the aging interval $t_a=12$, the proposed CP-Net reduces the NMSE by 28.6\% and 45.2\% relative to the LSTM--GRU and Kalman predictor, respectively, for ATA channels, the corresponding reductions for GTG channels are 44.1\% and 54.3\%, respectively. In addition, compared with the variant without channel-quality awareness (non-CQ-aware), the proposed CP-Net yields a slightly higher average NMSE. This is because the channel-quality-weighted loss places greater emphasis on weak links during training, thereby biasing the model toward more difficult samples and slightly sacrificing the global average performance. In Table~\ref{tab:nmse_2x6}, we compare CP-Net with CQ-aware against its non-aware variant by stratifying links into the top 10\% and bottom 10\% link-quality groups based on the average channel power. The results reveal a slight performance degradation on high-quality links but a pronounced improvement on weak links. For AG channels, the CQ-aware design leads to a 26.8\% NMSE increase for the top-10\% group, while achieving a 31.5\% NMSE reduction for the bottom-10\% group. Therefore, although the CQ-aware strategy slightly compromises the prediction accuracy of strong links, it substantially enhances the prediction quality of weak links.

\begin{figure}[t]
	\centering
	\includegraphics[width=0.8\linewidth]{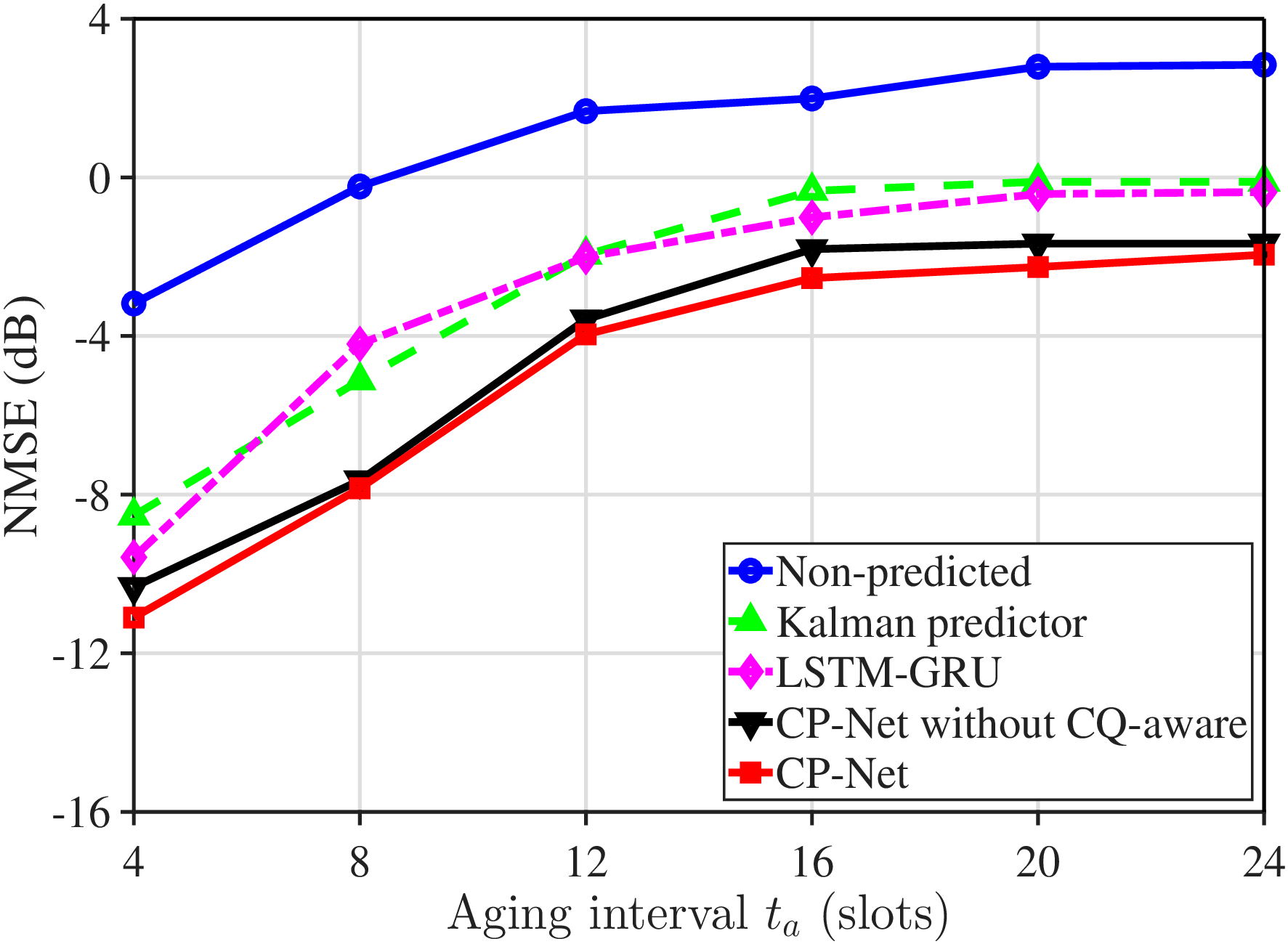}
	\caption{NMSE performance comparison versus aging interval for different methods under AG LoS/NLoS probabilistic switching.}
	\label{oddcp}
    
\end{figure}

\begin{figure}[t]
	\centering
	\includegraphics[width=0.8\linewidth]{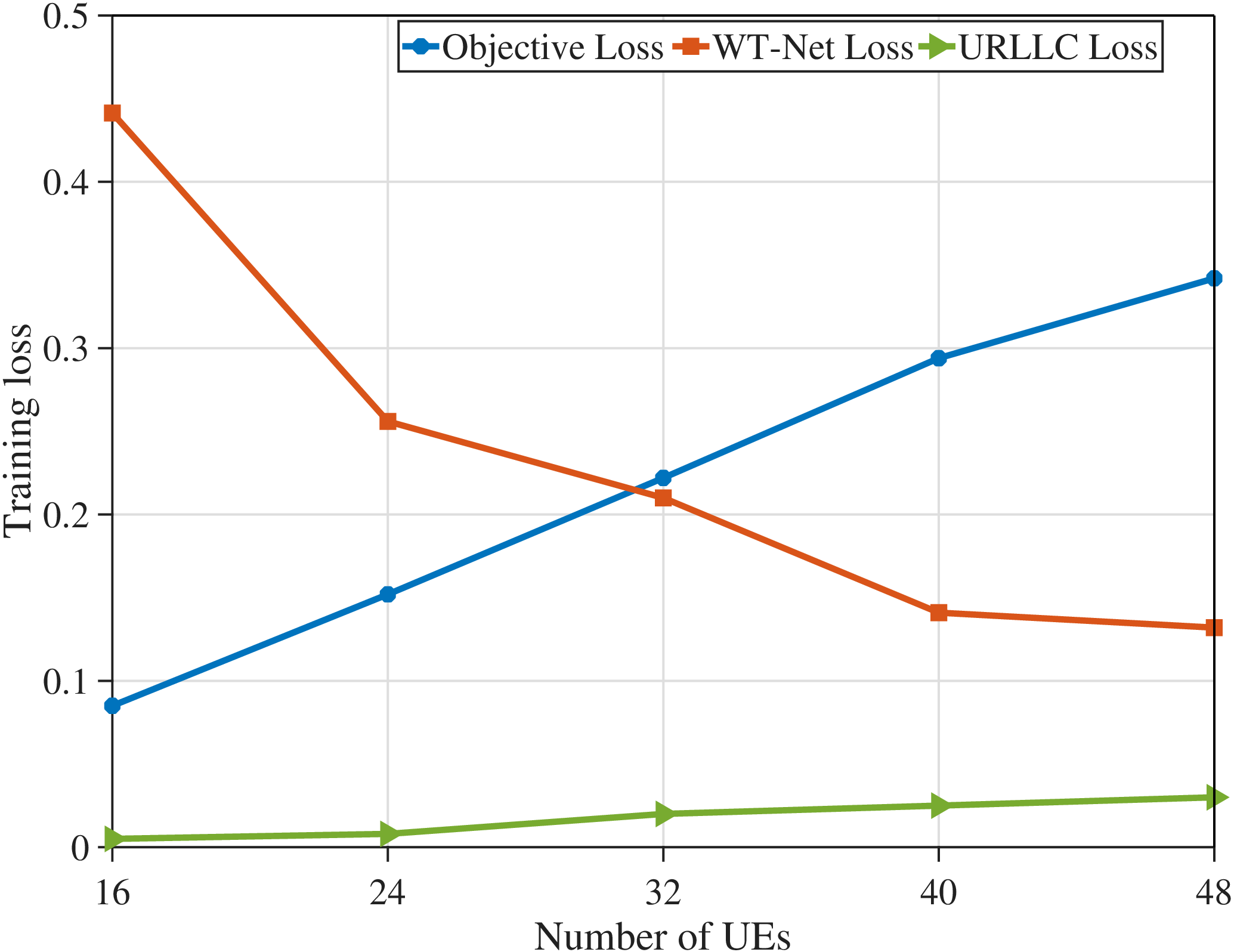}
	\caption{Training loss components versus the number of UEs.}
	\label{loss}
    
\end{figure}

\begin{table}[t]
	\centering
	\caption{NMSE comparison (In dB) for bottom 10\% and top 10\% channel-quality groups.}
	\label{tab:nmse_2x6}
	\scriptsize
	\setlength{\tabcolsep}{4pt}
	
	\begin{tabular}{l cc cc cc}
		\toprule
		\multirow{2}{*}{Quality group}
		& \multicolumn{2}{c}{ATA channels} & \multicolumn{2}{c}{AG channels} & \multicolumn{2}{c}{GTG channels} \\
		\cmidrule(lr){2-3}\cmidrule(lr){4-5}\cmidrule(lr){6-7}
		& CQ-aware & non-CQ & CQ-aware & non-CQ & CQ-aware & non-CQ \\
		\midrule
		Bottom 10\% & -13.74 & -10.52 & -6.26 & -4.62 & -3.16 & -2.47 \\
		Top 10\%    & -17.05 & -17.53 & -11.17 & -12.20 & -8.84 & -9.33 \\
		\bottomrule
	\end{tabular}
	\vspace{0.5mm}
\end{table}

To evaluate the generalization capability of the proposed method beyond the training distribution, we generate an out-of-distribution (OOD) test set with $1000$ independently randomized scenario samples, where the AG channels follow a 3GPP-inspired LoS/NLoS switching model. Specifically, according to \cite{9419751}, the time-varying LoS probability is denoted by $P^{\mathrm{3GPP}}_{\mathrm{LoS}}[t]\left(d_{\mathrm{2D}}[t],h_R\right)$, where $d_{\mathrm{2D}}[t]$ and $h_R$ denote the horizontal distance and aerial-UE altitude, respectively. Accordingly, the small-scale fading follows Rician fading with $K^{(\mathrm{h})}_{k,m}=10$ under LoS and Rayleigh fading with $K^{(\mathrm{h})}_{k,m}=0$ under NLoS. Fig.~\ref{oddcp} compares the NMSE performance of different methods on the OOD test set. Compared with the results in Fig.~3b, all methods exhibit a noticeable NMSE increase due to the distribution mismatch. Nevertheless, the proposed method still achieves the best prediction accuracy among all compared schemes, which provides evidence of its robustness under practical channel variations.

\subsection{Validation of Power Allocation}

\par Fig.~\ref{loss} illustrates the variation of the training loss versus the number of UEs $K$. The objective loss increases with $K$ due to the rise in IUI, which reduces the SINR experienced by the UEs and consequently degrades the optimization performance. The WT-Net loss decreases as $K$ increases, indicating the progressive specialization of experts within the MoE architecture. As system complexity grows, the weighting network allocates tasks more effectively to the most suitable experts, enabling each expert to focus on specific user types or channel conditions and thereby enhancing overall system efficiency. Meanwhile, the URLLC loss shows a slight increase with $K$, reflecting that higher IUI imposes greater challenges in meeting stringent reliability and latency requirements.

\begin{figure*}[t]
	\centering
	\subfloat[EE comparison\label{6a}]{
		\includegraphics[width=0.30\textwidth]{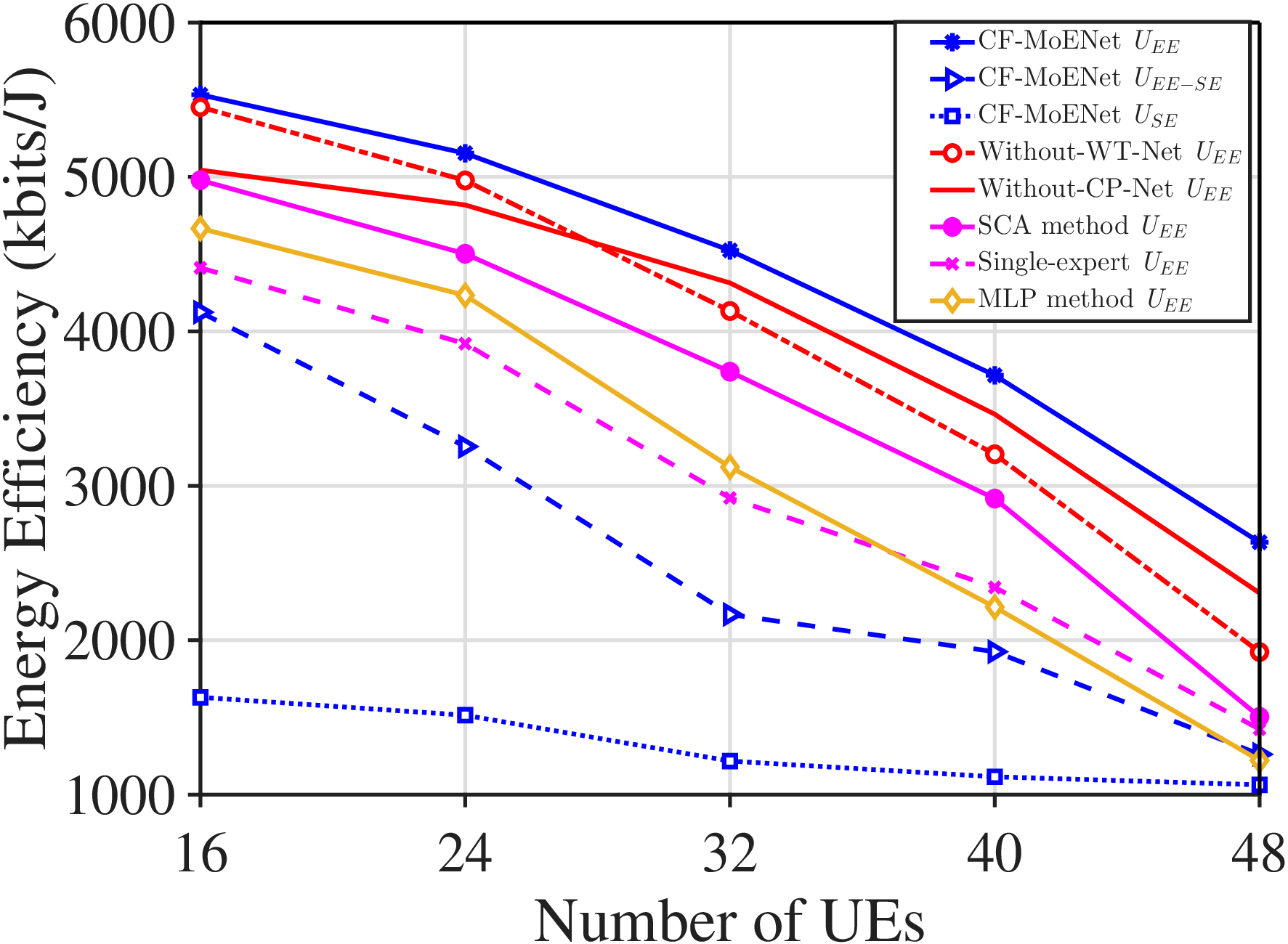}
	}
	\hfill
	\subfloat[SE comparison\label{6b}]{
		\includegraphics[width=0.30\textwidth]{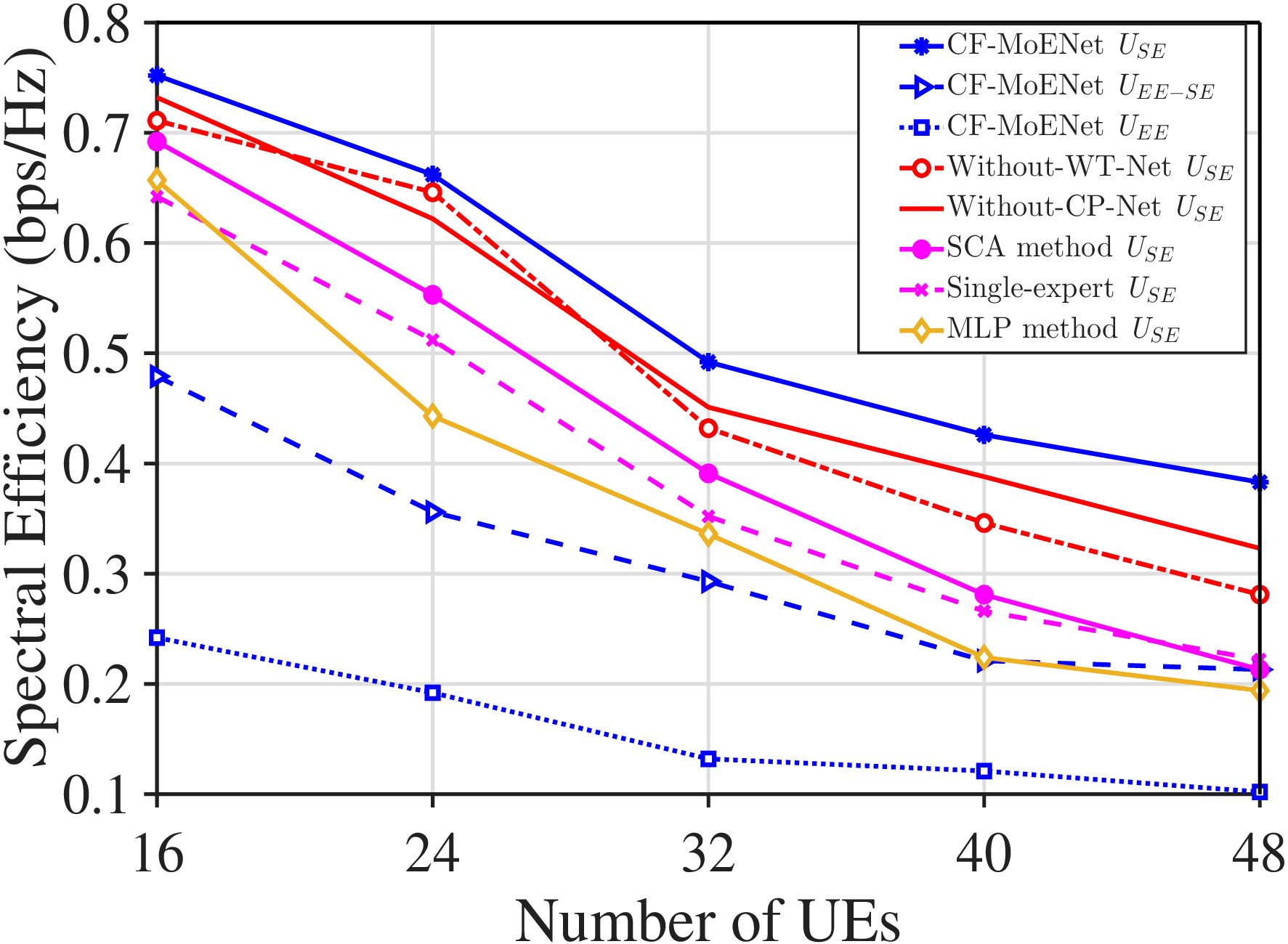}
	}
	\hfill
	\subfloat[USP comparison\label{6c}]{
		\includegraphics[width=0.30\textwidth]{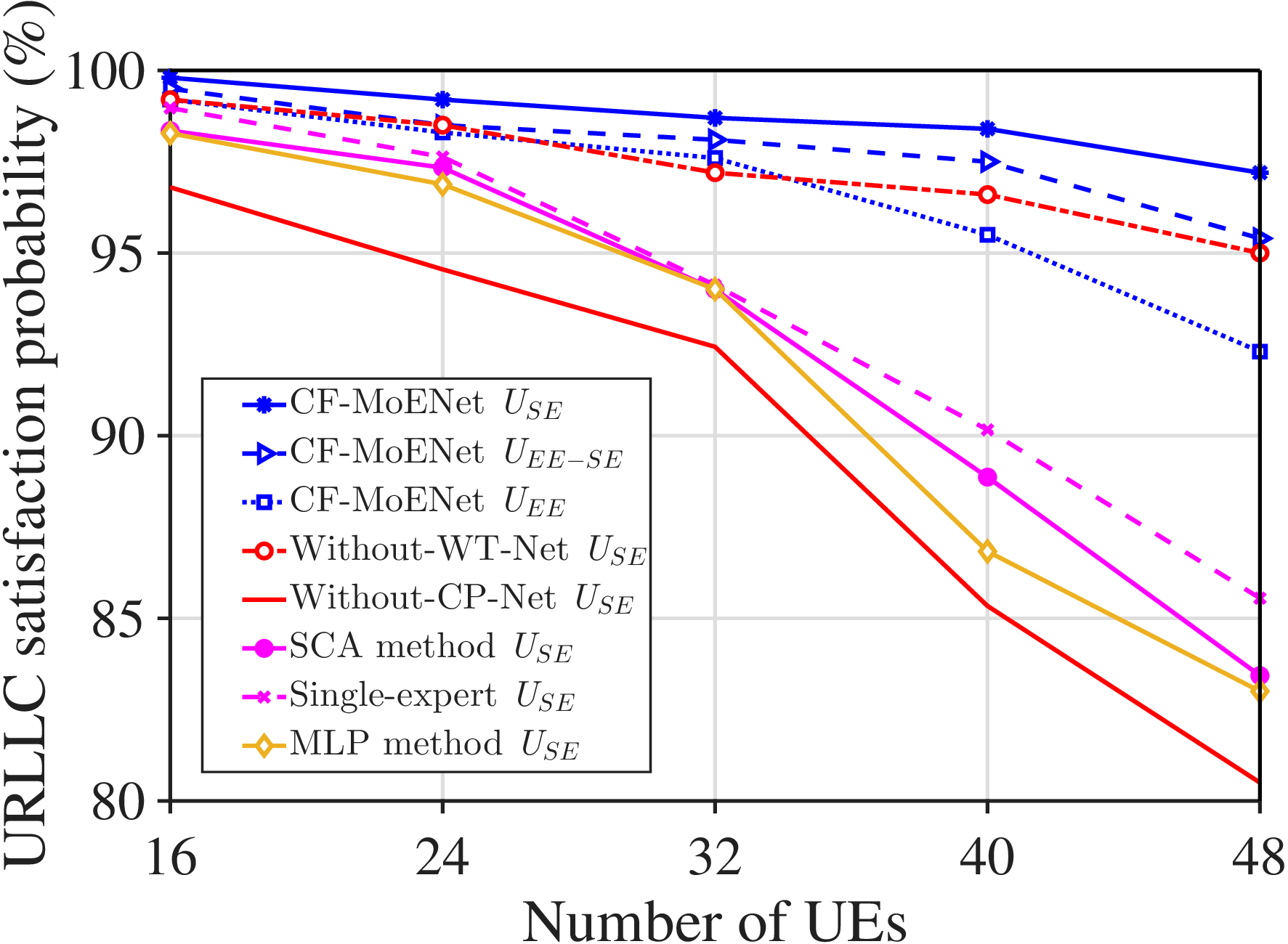}
	}
	\caption{Performance comparison of different power allocation strategies. (a) EE comparison, (b) SE comparison, and (c) USP comparison.}
	\label{fig6}
\end{figure*}

\par Figs. \ref{6a} and \ref{6b} compare the uplink EE and SE performance achieved by different power-allocation strategies, where “$U_{\mathrm{EE}}$”, “$U_{\mathrm{SE}}$”, and “$U_{\mathrm{EE\text{-}SE}}$” denote EE-oriented, SE-oriented, and EE--SE trade-off UEs, respectively. As shown in Fig.~\ref{6a}, the EE of all schemes decreases as the number of UEs $K$ increases, mainly due to stronger IUI. We further observe consistent EE degradation when removing CP-Net or WT-Net, or when replacing the MoE architecture with a single-expert network. Specifically, at $K=32$, the EE of EE-oriented UEs decreases by approximately 4.66\% without CP-Net, 8.6\% without WT-Net, and 35.43\% with the single-expert variant, relative to the full model. These ablation results indicate that CP-Net mitigates channel aging and improves CSI quality, thereby enabling more effective power-control decisions, while the WT-Net–assisted multi-expert architecture adaptively fuses expert outputs and yields more robust and generalizable power allocation under heterogeneous service demands and channel conditions. Furthermore, SE-oriented UEs achieve lower EE than EE-oriented UEs since higher data rates typically require higher transmit power, whereas the EE--SE trade-off UEs attain intermediate performance. Compared with the baseline methods, the proposed approach provides substantial gains. For instance, at $K=16$, it improves the EE of EE-oriented UEs by 17.36\% and 31\% over the SCA and MLP baselines, respectively. As shown in Fig.~\ref{6b}, consistent with the EE results, the proposed CF-MoENet achieves significantly higher SE than all comparison schemes. Moreover, relative to the baseline algorithms, the proposed MoE-based method improves the SE by about 20.53\% and 31.7\% over SCA and MLP algorithms, respectively. These results collectively indicate the effectiveness and robustness of the proposed learning-based framework, outperforming both baselines under the considered settings.

\begin{table}[t]
\centering
\caption{USP comparison between the CQ-aware and non-CQ-aware schemes under different numbers of UEs.}
\label{uspnon}
\renewcommand{\arraystretch}{1.1}
\setlength{\tabcolsep}{5pt}
\footnotesize
\begin{tabular}{l|ccccc}
\toprule
\textbf{Method} & \textbf{$K=8$} & \textbf{$K=16$} & \textbf{$K=24$} & \textbf{$K=32$} & \textbf{$K=40$} \\
\midrule
CF-MoENet     & 99.81 & 99.24 & 98.77 & 98.40 & 97.22 \\
non-CQ-aware     & 99.23 & 98.52 & 98.30 & 97.82 & 96.63 \\

\bottomrule
\end{tabular}
\vspace{1mm}
\end{table}

\par In Fig.~\ref{6c}, we adopt the URLLC satisfaction probability (USP) \cite{9286851} to characterize the reliability performance of the system. Specifically, given the payload size $D$ and the transmission duration $t_{\mathrm{tr}}$, the target operating rate is defined as $R_b \triangleq \frac{D}{B t_{\mathrm{tr}}}$. Accordingly, the decoding EP of UE $k$ is evaluated at the operating rate $R_b$ based on \eqref{eq13}, which is denoted by $\epsilon_k^{\star}[t]$. The USP at sampling instant $t$ is then defined as
\begin{equation}
\small
\tag{26}
\label{eq:usr_joint_eps}
USP_{\rm joint}[t]
=\frac{1}{K}\sum_{k=1}^{K}\mathrm{H}\!\Big(\epsilon_b-\epsilon_k^{\star}[t]\Big),
\end{equation}

\noindent where $\mathrm{H}(\cdot)$ denotes the Heaviside step function, which equals $1$ if the argument is non-negative and $0$ otherwise.

\begin{table*}[t]
\centering
\caption{Pilot time--frequency resource sensitivity: EE, SE, and USP versus $\tau_{\mathrm{p}}$.}
\label{pilot}
\footnotesize
\setlength{\tabcolsep}{3.5pt}
\renewcommand{\arraystretch}{1.12}
\begin{tabular}{cccccccccc}
\toprule
\multirow{2}{*}{$\tau_{\mathrm{p}}$ (CU)}
& \multicolumn{3}{c}{\textbf{EE (kbits/J)}}
& \multicolumn{3}{c}{\textbf{SE (bps/Hz)}}
& \multicolumn{3}{c}{\textbf{USP (\%)}} \\
\cmidrule(lr){2-4}\cmidrule(lr){5-7}\cmidrule(lr){8-10}
& $\text{CF-MoE}_{U_{\mathrm{EE}}}$ & $\text{SCA}_{U_{\mathrm{EE}}}$ & $\text{MLP}_{U_{\mathrm{EE}}}$
& $\text{CF-MoE}_{U_{\mathrm{SE}}}$ & $\text{SCA}_{U_{\mathrm{SE}}}$ & $\text{MLP}_{U_{\mathrm{SE}}}$
& $\text{CF-MoE}_{U_{\mathrm{SE}}}$ & $\text{SCA}_{U_{\mathrm{SE}}}$ & $\text{MLP}_{U_{\mathrm{SE}}}$ \\
\midrule
500 & 5122.47 & 4519.81 & 4414.48 & 0.674 & 0.550 & 0.589 & 99.13 & 95.59 & 95.95 \\
1000 & 5256.79 & 4731.06 & 4521.22 & 0.688 & 0.637 & 0.608 & 99.47 & 97.52 & 96.88 \\
1500 & 5153.45 & 4502.55 & 4233.42 & 0.662 & 0.603 & 0.583 & 99.24 & 97.34 & 96.74 \\
2000 & 4834.20 & 4123.84 & 3880.17 & 0.629 & 0.552 & 0.528 & 99.21 & 97.14 & 96.43 \\
2500 & 4510.99 & 3967.52 & 3663.25 & 0.601 & 0.528 & 0.467 & 98.71 & 96.55 & 96.20 \\
\bottomrule
\end{tabular}
\end{table*}

\begin{figure}[t]
	\centering
	\includegraphics[width=0.8\linewidth]{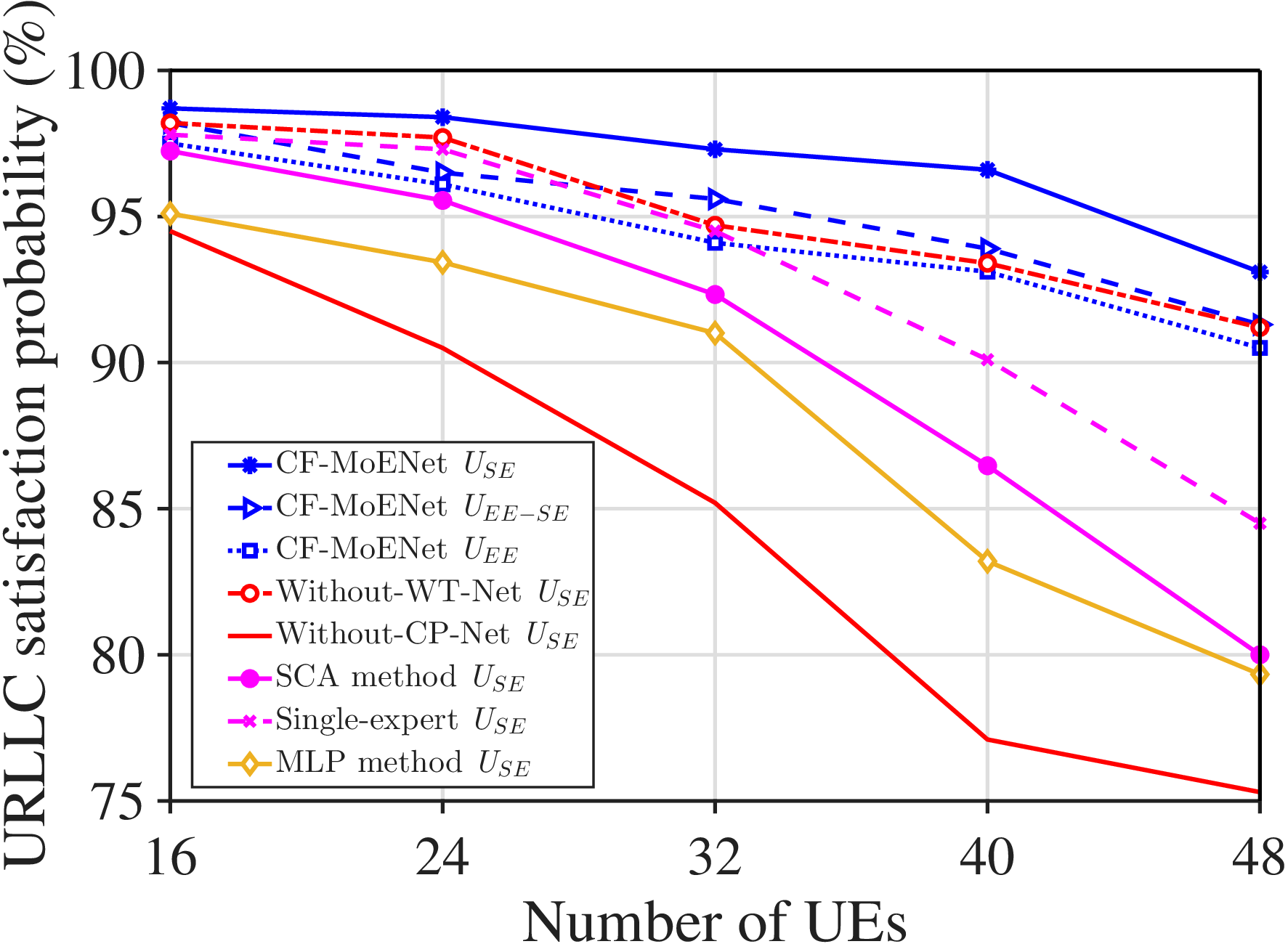}
	\caption{Comparison of USP across different power allocation strategies for the bottom 10\% UEs ranked by SINR.}
	\label{fig7}
\end{figure}
\begin{table*}[t]
\centering
\caption{Transmission latency budget sensitivity: EE, SE, and USP versus $t_{\mathrm{tr}}$.}
\label{tab:latency}
\footnotesize
\setlength{\tabcolsep}{3.5pt}
\renewcommand{\arraystretch}{1.12}
\begin{tabular}{cccccccccc}
\toprule
\multirow{2}{*}{$t_{\mathrm{tr}}$ (ms)}
& \multicolumn{3}{c}{\textbf{EE (kbits/J)}}
& \multicolumn{3}{c}{\textbf{SE (bps/Hz)}}
& \multicolumn{3}{c}{\textbf{USP (\%)}} \\
\cmidrule(lr){2-4}\cmidrule(lr){5-7}\cmidrule(lr){8-10}
& $\text{CF-MoE}_{U_{\mathrm{EE}}}$ & $\text{SCA}_{U_{\mathrm{EE}}}$ & $\text{MLP}_{U_{\mathrm{EE}}}$
& $\text{CF-MoE}_{U_{\mathrm{SE}}}$ & $\text{SCA}_{U_{\mathrm{SE}}}$ & $\text{MLP}_{U_{\mathrm{SE}}}$
& $\text{CF-MoE}_{U_{\mathrm{SE}}}$ & $\text{SCA}_{U_{\mathrm{SE}}}$ & $\text{MLP}_{U_{\mathrm{SE}}}$ \\
\midrule
0.25 & 4416.14 & 3779.50 & 3308.58 & 0.569 & 0.523 & 0.446 & 94.99 & 87.43 & 82.54 \\
0.30 & 4633.58 & 3947.79 & 3642.59 & 0.597 & 0.551 & 0.491 & 97.21 & 92.37 & 88.52 \\
0.35 & 4796.01 & 4299.16 & 3885.14 & 0.622 & 0.583 & 0.535 & 98.63 & 94.59 & 92.10 \\
0.40 & 5153.45 & 4502.55 & 4233.42 & 0.662 & 0.603 & 0.583 & 99.24 & 97.34 & 96.74 \\
0.45 & 5224.13 & 4668.29 & 4420.13 & 0.683 & 0.628 & 0.614 & 99.67 & 98.36 & 97.13 \\
\bottomrule
\end{tabular}
\end{table*}

\par As the number of UEs increases, the USP decreases, mainly due to the intensified IUI and heavier network load, both of which reduce the achievable data rates. Moreover, SE-oriented UEs achieve better URLLC performance than EE-oriented and trade-off UEs, since energy-efficient transmissions usually operate at lower power levels, which limits the achievable rates and consequently degrades reliability. At $K=32$, for SE-oriented UEs, the full CP-Net–enabled MoE-Net achieves about 1.51\%, 6.35\%, and 4.6\% higher USP than the variants without CP-Net, without WT-Net, and with a single-expert network, respectively. Moreover, it outperforms the SCA and MLP baselines by about 4.3\% and 5\%, respectively. From Table~V, it can be observed that the proposed CF-MoENet with the CQ-aware design consistently achieves higher USP than the non-CQ-aware variant under different numbers of UEs. This result indicates that incorporating channel-quality information is beneficial for improving URLLC service satisfaction. We further define the USP for the bottom 10\% UEs ranked by SINR at instant $t$ as

\begin{figure}[t]
	\centering
	\includegraphics[width=0.8\linewidth]{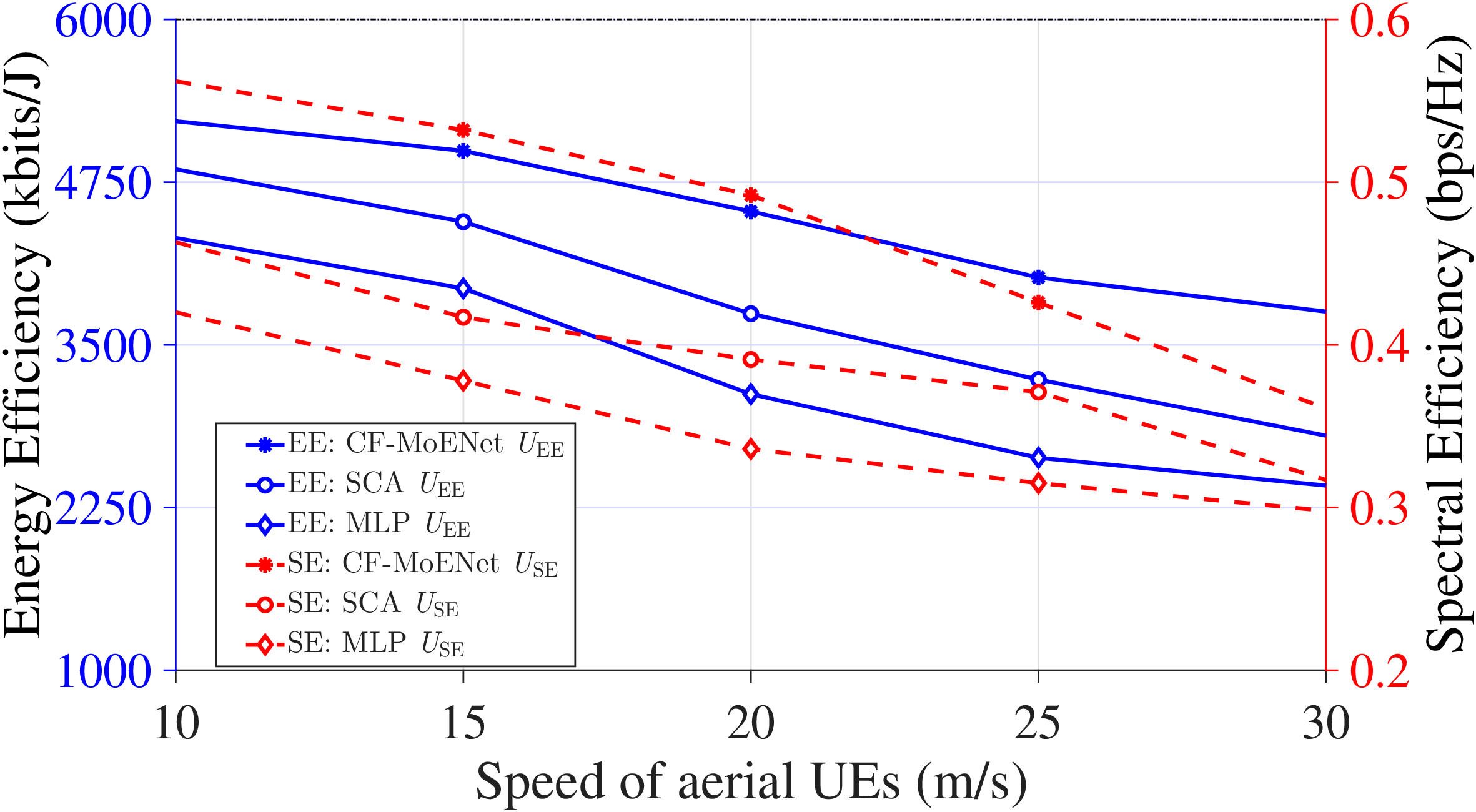}
	\caption{Comparison of EE and SE versus different aerial-UE speeds.}
	\label{fig8}
\end{figure}

\begin{figure}[t]
	\centering
	\includegraphics[width=0.8\linewidth]{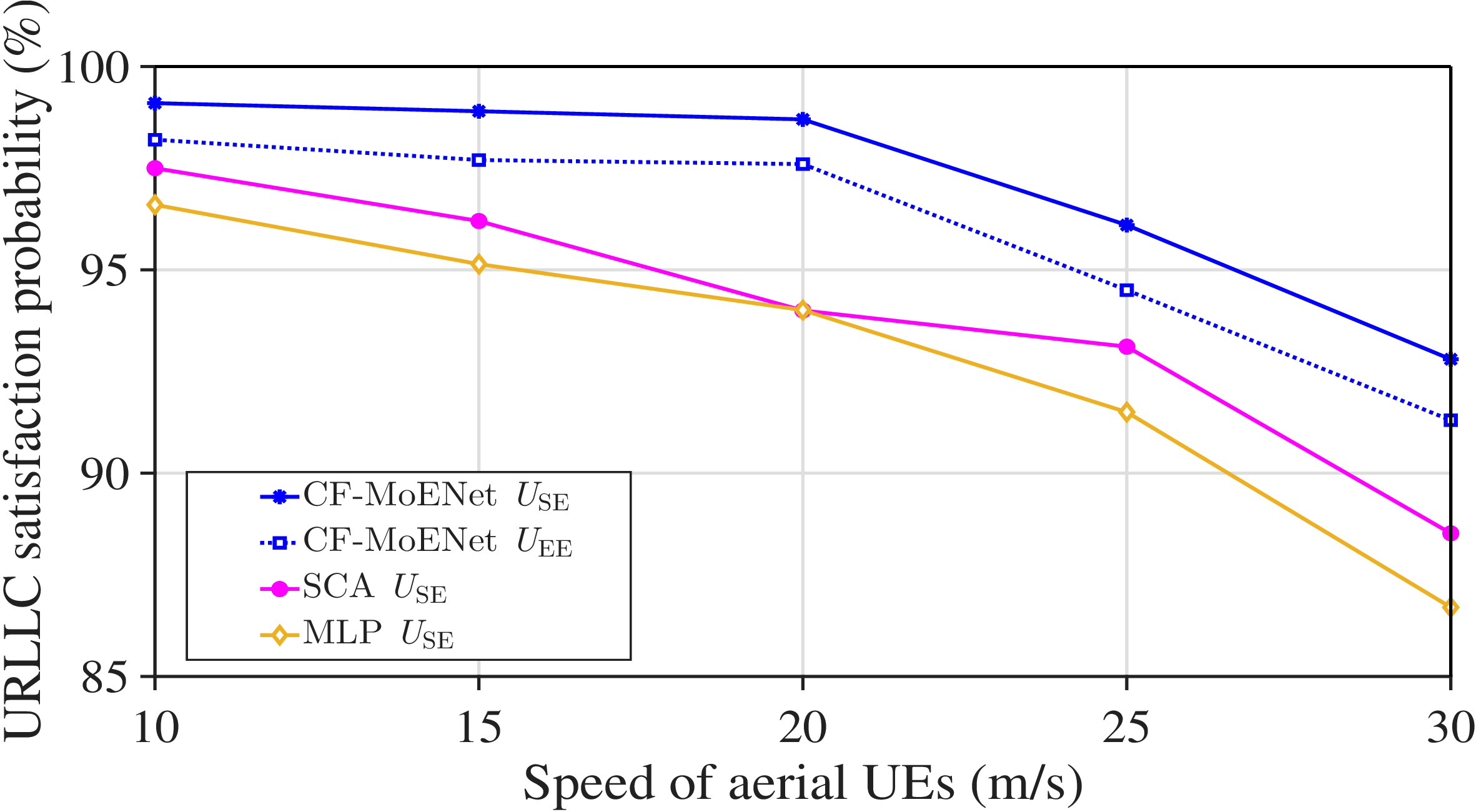}
	\caption{Comparison of USP versus different aerial-UE speeds.}
	\label{fig9}
\end{figure}
\begin{figure}[t]
	\centering
	\includegraphics[width=0.8\linewidth]{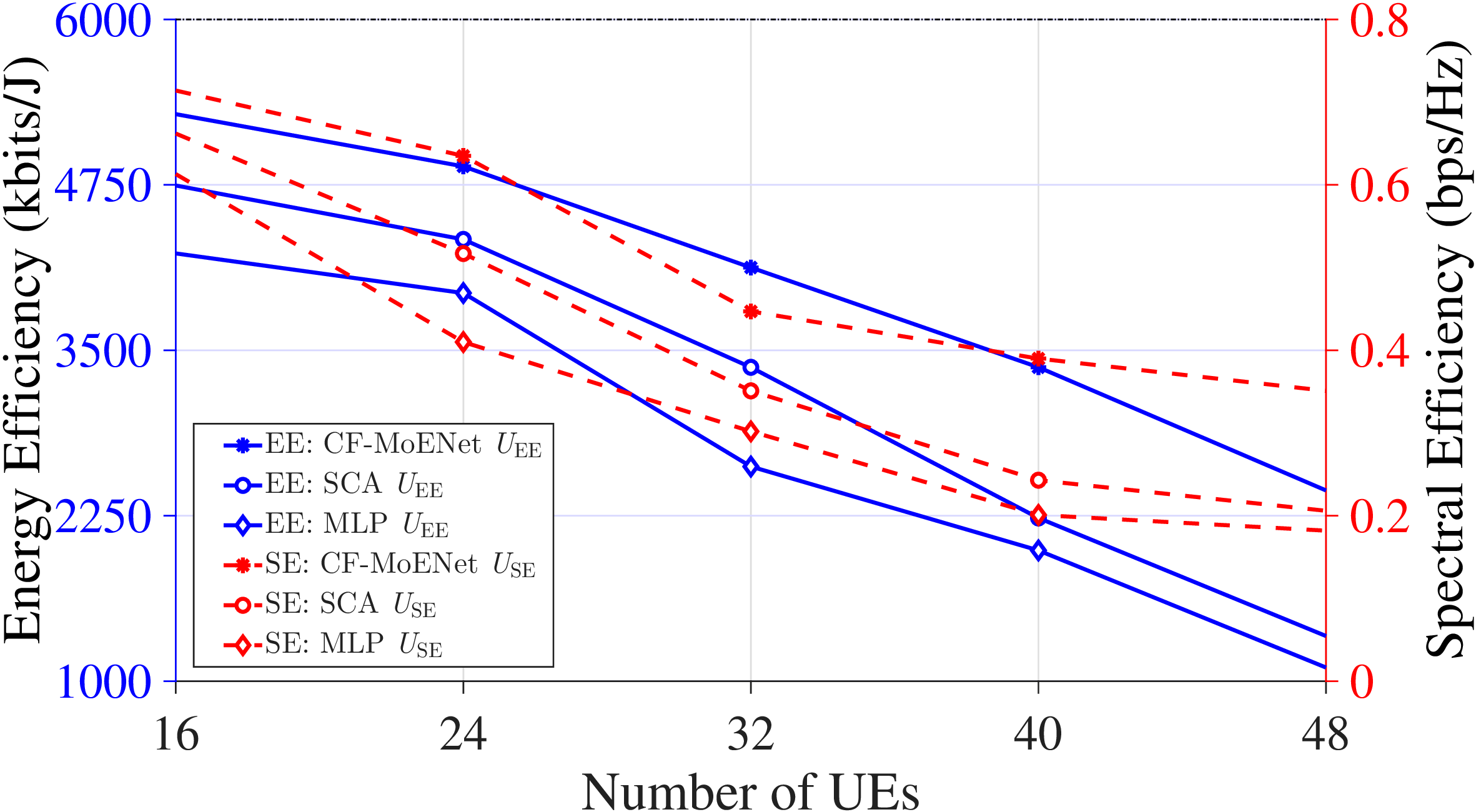}
	\caption{EE and SE Comparison under AG LoS/NLoS probabilistic switching.}
	\label{fig10}
\end{figure}
\begin{figure}[t]
	\centering
	\includegraphics[width=0.8\linewidth]{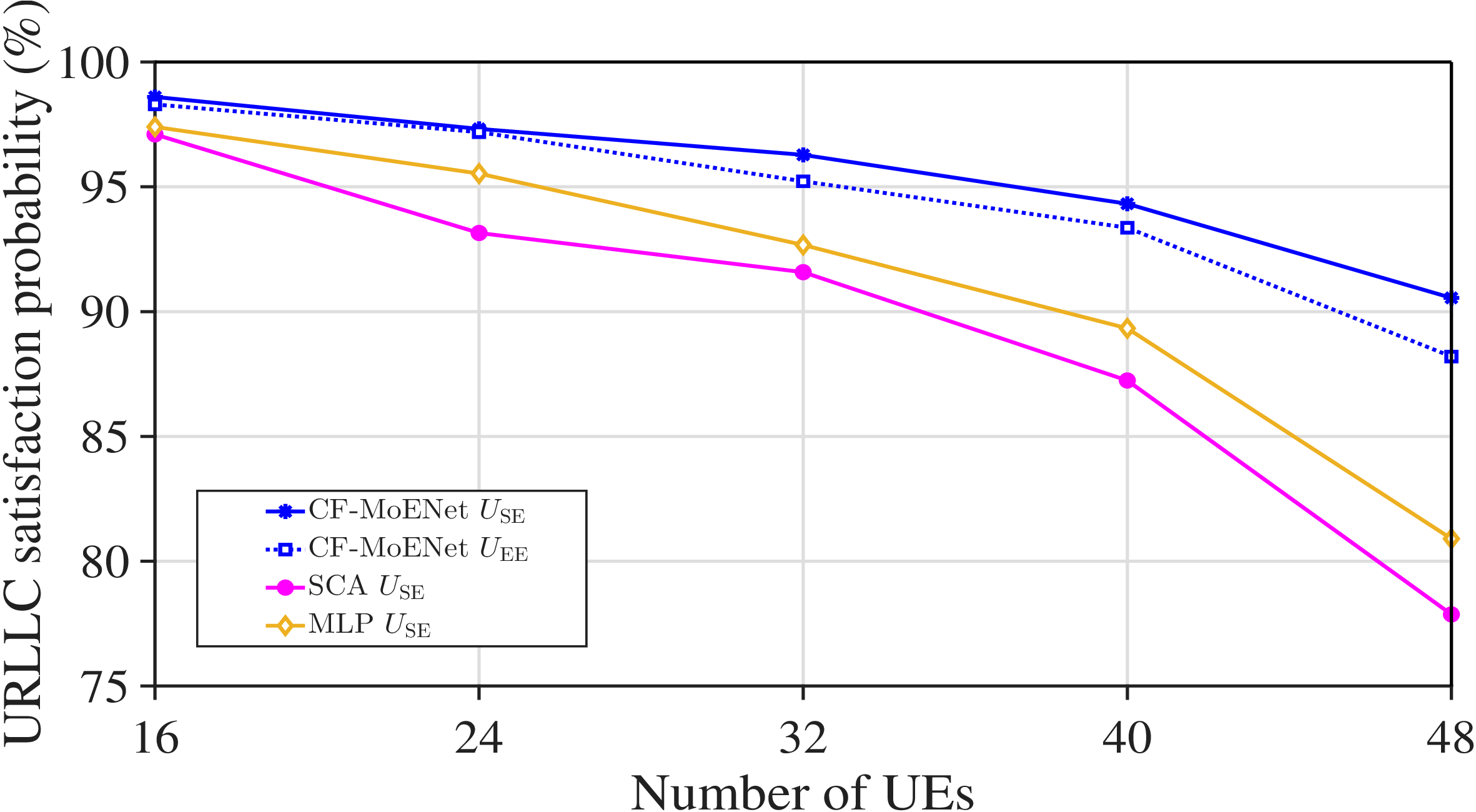}
	\caption{USP Comparison under AG LoS/NLoS probabilistic switching.}
	\label{fig11}
\end{figure}
\begin{equation}
\small
\tag{27}
\label{eq:usr_weak}
USP^{10\%}_{\rm joint}[t]
=\frac{1}{|\mathcal{K}_{\rm weak}|}\sum_{k\in\mathcal{K}_{\rm weak}}
\mathrm{H}\!\Big(\epsilon_b-\epsilon_k^{\star}[t]\Big),
\end{equation}

\noindent where $|\mathcal{K}_{\rm weak}|$ denotes the cardinality of $\mathcal{K}_{\rm weak}$, and $\mathcal{K}_{\rm weak}$ denotes the set of the bottom 10\% UEs ranked according to their effective SINR values defined in (21). As shown in Fig.~\ref{fig7}, the bottom 10\% SINR UEs achieve lower performance than the average across all UEs, as low-SINR UEs are inherently more challenging to optimize. Even so, the proposed CF-MoENet consistently outperforms the benchmark methods for this bottom group, indicating that the achieved overall performance gains do not come at the cost of weak-user reliability.

\par To examine the performance of the proposed network under different parameter settings, we compare the EE, SE, and USP metrics under different pilot blocklengths $\tau_{\mathrm{p}}$ and transmission-latency budgets $t_{\mathrm{tr}}$. Table~\ref{pilot} indicates that the impact of the pilot blocklength $\tau_{\mathrm{p}}$ is not monotonic. Specifically, as $\tau_{\mathrm{p}}$ increases from a relatively small value, the EE, SE, and USP metrics first improve because a larger pilot resource budget enhances channel-estimation accuracy and alleviates CSI uncertainty, which in turn benefits the subsequent transmission and optimization stages. However, when $\tau_{\mathrm{p}}$ becomes too large, the remaining data blocklength $\tau_{\mathrm{s}} = N - \tau_{\mathrm{p}}$ is significantly reduced. Under the FBL regime, the reduction in payload resources gradually outweighs the estimation gain brought by longer pilots, thereby causing performance degradation. As mentioned above, since $t_{\mathrm{oh}}$ is deployment-dependent, the available $t_{\mathrm{tr}}$ generally differs across practical scenarios. Table~\ref{tab:latency} reports the EE, SE, and USP metrics for different transmission-time budgets $t_{\mathrm{tr}}$ under a fixed URLLC E2E latency constraint $t_{b}$. As $t_{\mathrm{tr}}$ increases, all metrics improve because a larger transmission-time budget provides more channel uses and alleviates the finite-blocklength penalty, thereby enabling more reliable and efficient transmission. Overall, the proposed method consistently outperforms the baselines under different $\tau_{\mathrm{p}}$ and $t_{\mathrm{tr}}$ settings, providing stronger empirical evidence of robustness to variations in pilot length and transmission-time budget within the considered setting.

\par In Figs.~\ref{fig8} and \ref{fig9}, we evaluate the performance under different aerial-UE speeds at $K=32$ to examine the robustness of the proposed method to mobility variations. As the speed increases, the performance of all schemes gradually degrades because larger Doppler shifts and more severe channel aging make CSI prediction more challenging. Nevertheless, the proposed CF-MoENet consistently outperforms the iterative SCA and lightweight MLP baselines in terms of EE, SE, and USP across the considered speed range. For instance, at $v=15$~m/s, CF-MoENet improves EE by about $10.91\%$ and $28.75\%$ over SCA and MLP, respectively. These results demonstrate that the proposed framework can effectively mitigate mobility-induced CSI degradation and maintain robust performance in high-mobility UAM scenarios. Moreover, Figs.~\ref{fig10} and \ref{fig11} compare the power-allocation performance on the OOD test set with probabilistic LoS/NLoS switching for AG links. Compared with the original test set, all schemes experience performance degradation, since LoS/NLoS switching increases channel non-stationarity and CSI uncertainty. Nevertheless, the proposed CF-MoENet still outperforms the considered benchmark methods, providing additional evidence of robustness under the tested AG propagation variation.

\subsection{Complexity Analysis}

\par In this subsection, we analyze the computational complexity of the considered methods. For channel prediction, the overall computational complexity is $\mathcal{O}\!\left(KMV^2L^2\right)$. The nonlinear Kalman predictor requires recursive updates of the state and covariance matrices for each UE--AP link, yielding complexity $\mathcal{O}\!\left(KM(2L)^3\right)$, while the LSTM--GRU predictor has complexity $\mathcal{O}\!\left(KMV(2Lh+h^2)\right)$, where $h$ denotes the hidden dimension. For power allocation, the expert networks have complexity on the order of $\mathcal{O}\!\left(KM^2L+KL^2\right)$, whereas WT-Net incurs complexity $\mathcal{O}\!\left(KM^2L\right)$. Hence, the overall online complexity of MoE-Net + WT-Net is dominated by the expert networks and scales as $\mathcal{O}\!\left(KM^2L+KL^2\right)$. Moreover, the SCA baseline requires iteratively solving a sequence of convexified subproblems, and its overall computational complexity scales as $\mathcal{O}\!\left(I_{\mathrm{SCA}}K^3\right)$, where $I_{\mathrm{SCA}}$ denotes the number of iterations. For the MLP baseline with a two-layer FC structure, the complexity is $\mathcal{O}\!\left(K(2MLH+H)\right)$, which reduces to $\mathcal{O}(KML)$ when the hidden dimension $H$ is treated as a constant.
\par Moreover, for channel prediction, the average per-link inference latencies of the proposed CP-Net, the nonlinear Kalman predictor, and the LSTM--GRU predictor are 0.325 ms, 6.23 ms, and 0.238 ms, respectively. For power allocation, the online decision latency under different numbers of UEs is reported in Table~\ref{tab:iter_latency}. Compared with the iterative SCA algorithm, the learning-based methods achieve significantly lower decision latency, since most of their computational burden is shifted to offline training, whereas the online stage only involves lightweight feedforward inference. For the iterative SCA baseline, the maximum number of iterations is set to 50, the SCA method requires 38 iterations on average when $K=32$, which leads to substantially higher online computational overhead and makes it less suitable for practical deployment under stringent URLLC latency requirements. Moreover, although CP-Net and MoE-Net are moderately more complex than lightweight baselines, they still maintain low and stable online latency while remaining substantially more efficient than iterative optimization methods. These results indicate a favorable practical tradeoff of the proposed framework for URLLC aerial--terrestrial CF-mMIMO systems.

\begin{table}[t]
\centering
\caption{Online decision latency comparison (in ms) under different numbers of UEs.}
\label{tab:iter_latency}
\renewcommand{\arraystretch}{1.1}
\setlength{\tabcolsep}{5pt}
\footnotesize
\begin{tabular}{l|ccccc}
\toprule
\textbf{Method} & \textbf{$K=8$} & \textbf{$K=16$} & \textbf{$K=24$} & \textbf{$K=32$} & \textbf{$K=40$} \\
\midrule
SCA     & 2550 & 3280 & 3990 & 5290 & 7530 \\
MLP     & 0.25 & 0.46 & 0.83 & 1.20 & 1.84 \\
MoE-Net+WT-Net & 0.48 & 0.92 & 1.42 & 2.09 & 2.70 \\
\bottomrule
\end{tabular}
\vspace{1mm}
\end{table}

\section{Conclusion}

In this paper, we proposed a hybrid aerial–terrestrial CF-mMIMO network to serve URLLC UEs in UAM scenarios and designed a deep learning framework for uplink optimization. Specifically, CP-Net predicts aged CSI to mitigate channel aging, while MoE-Net with WT-Net adaptively allocates power to optimize EE, SE, and their trade-off. Simulation results demonstrate that the proposed approach significantly improves channel prediction accuracy while effectively balancing the performance objectives of heterogeneous UEs. Compared with baseline methods, the proposed framework demonstrates superior generalization and robustness, while striking an effective balance between performance and computational complexity, making it a promising solution for diverse low-altitude UAM applications under URLLC constraints. Overall, the current work focuses on transmission-oriented optimization under the adopted aerial--terrestrial CF-mMIMO systems, while several practical deployment aspects related to system implementation, hardware realism, and model adaptation still warrant further investigation. Future work will explicitly incorporate processing and fronthaul latency and explore overhead-aware and hardware-robust design, online adaptation, data-efficient continual model updating under long-term environmental variations, and more principled multi-objective optimization frameworks for practical aerial--terrestrial CF-mMIMO deployments.

\bibliography{refer.bib}

\end{document}